\documentclass[pra,twocolumn,groupedaddress,showpacs,floatfix]{revtex4}

\usepackage{graphics}
\usepackage{graphicx}
\usepackage{bm}
\usepackage{amsmath}
\usepackage{amsfonts} 
\usepackage{amssymb}
\usepackage{latexsym}
\usepackage{color}
\usepackage{ctable}

\newcommand{\bra}[1]{\langle #1 |}
\newcommand{\ket}[1]{| #1 \rangle}

\begin{document}

\title{Extending the validity range of quantum optical master equations}
\author{Adam Stokes, Andreas Kurcz, Tim P. Spiller, and Almut Beige}
\affiliation{The School of Physics and Astronomy, University of Leeds, Leeds, LS2 9JT, United Kingdom} 

\date{\today}

\begin{abstract}
This paper derives master equations for an atomic two-level system for a large set of unitarily equivalent Hamiltonians without employing the rotating wave and certain Markovian approximations. Each Hamiltonian refers to physically different components as representing the ``atom" and as representing the ``field" and hence results in a different master equation, when assuming a photon-absorbing environment. It is shown that the master equations associated with the minimal coupling and the multipolar Hamiltonians predict enormous stationary state narrowband photon emission rates, even in the absence of external driving, for current experiments with single quantum dots and colour centers in diamond. These seem to confirm that the rotating wave Hamiltonian identifies the components of the atom-field system most accurately.
\end{abstract}
\pacs{03.65.Yz, 31.30.J-}

\maketitle

\section{Introduction} \label{Introduction}

Due to fast experimental progress \cite{progress,progress2,Atature,Jelezko}, it becomes increasingly important to model quantum optical systems with spontaneous photon emission more and more accurately \cite{Andreas4,Andreas,Andreas3}. Motivated by this, we revisit the derivation of the master equation of an atomic two-level system, like a single quantum dot, a single colour centre in diamond, or a well-localised atom or ion, within a free radiation field without employing the rotating wave approximation (RWA) and certain Markovian approximations. The starting point of our derivation is, as usual, a quantum electrodynamic Hamiltonian of the form 
\begin{eqnarray}\label{h1}
H &=& H_{\rm A} + H_{\rm F} + H_{\rm AF} \, , 
\end{eqnarray}
where $H_{\rm A}$, $H_{\rm F}$ and $H_{\rm AF}$ are the atom, the field, and the atom-field interaction components. This Hamiltonian acts on a Hilbert space ${\mathcal H} = {\mathcal H}_{\rm A} \otimes {\mathcal H}_{\rm F}$ whose subspaces ${\mathcal H}_{\rm A}$ and ${\mathcal H}_{\rm F}$ are the eigenspaces of $H_{\rm A}$ and $H_{\rm F}$, respectively. The purpose of the master equation is to summarise the effective time evolution of the atomic system induced by this Hamiltonian and by the coupling of the field to a photon-absorbing environment.

But there is a problem: There are in principle infinitely many unitarily equivalent representations $H'$ of the Hamiltonian $H$ in Eq.~(\ref{h1}). Each of these Hamiltonians relates to $H$ via a unitary transformation $R$, 
\begin{eqnarray}\label{h2}
H' &=& R \, H \, R^\dagger \, . 
\end{eqnarray}
Now, suppose $R$ does not commute with $H_{\rm A}$ and $H_{\rm F}$. Then, although physically equivalent, $H$ and $H'$ correspond to different decompositions of the Hilbert space ${\mathcal H}$ into subspaces ${\mathcal H}_{\rm A}$ and ${\mathcal H}_{\rm F}$. Both refer in general to physically different components as representing the ``system" and as representing the ``field" \cite{drummond}. The Hamiltonians $H$ and $H'$ therefore result in general in different master equations. This problem only vanishes when certain Markovian approximations are employed which reduce every master equation to the standard Born-Markov form -- independent of the chosen representation of $H$. When going beyond these approximations, we are faced with the question as to which representation of $H$ should be employed?

That there are many different representations of $H$ is directly related to the fact that there are many possible choices of gauge of the electromagnetic field. The most prominent form of $H$ is the so-called {\em minimal coupling} Hamiltonian which is obtained after applying the canonical quantisation procedure to the classical Lagrangian in Coulomb gauge \cite{craig}. Starting from this Hamiltonian, a more general form of the atom-field Hamiltonian, including the {\em multipolar} Hamiltonian, is obtained through the application of the Power-Zienau-Woolley (PZW) transformation \cite{cohen1,woolley1,woolley2}. Further representations of $H$ are obtained through a generalised PZW transformation which depends on a set of parameters $\alpha_k$. As pointed out by Drummond \cite{drummond}, a particular choice of $\alpha_k$'s can be made to eliminate the counter-rotating terms in the atom-field interaction $H_{\rm AF}$, when assuming that the size of the atomic system is much smaller than a typical optical wave length (electric dipole approximation). In the following we refer to the resulting Hamiltonian as the {\em rotating wave} Hamiltonian, since almost the same Hamiltonian, ie.~a Hamiltonian of the same form but with different coupling constants, arises when applying the RWA. 

This paper considers the vast set of unitarily equivalent atom-field Hamiltonians generated by the above mentioned generalisation of the PZW transformation \cite{drummond} and derives a master equation for every one of them. It is then shown that most of these equations predict enormous stationary state photon emission rates $I_{\rm ss}$, even in the absence of external driving, when we consider the parameters of recent experiments with single quantum dots \cite{Atature} and single colour centers in diamond \cite{Jelezko}. The only master equation which yields $I_{\rm ss} \equiv 0$ is the one associated with the rotating wave Hamiltonian. This means, this Hamiltonian (and not the minimal coupling or the multipolar Hamiltonians) constitutes the physically most relevant representation of the atom-field Hamiltonian in Eq.~(\ref{h1}). As we shall see below, the corresponding master equation is essentially the same as the standard Born-Markov master equation. However, it is no longer the result of approximations but the result of a specific choice of $R$ which identifies the bare states of the atomic system and its free radiation field correctly.

Many different ways of deriving master equations, ie.~of deriving an effective equation for the time evolution of the atomic system, can be found in the literature. Due to a great variety of microscopic models, different authors usually apply different approximations. For example, many derivations of master equations rely on the so-called Born approximation \cite{Agarwal,noise,Mandel,Walls,Cohen2,Breuer}. This approximation assumes that the state of a system and its bath can be written as a product state of the form $\rho_{\rm A} \otimes \rho_{\rm F}$ at any possible initial time $t$, where $\rho_{\rm F}$ denotes the thermal equilibrium of the bath. In general it is emphasized that the Born approximation applies well only in certain situations, like the weak coupling regime, where memory effects remain negligible \cite{Davies,Petro}.   

Moreover, many authors emphasize that there is a strong overlap between the validity range of the Born approximation and the validity range of certain Markovian approximations. A system is Markovian, ie.~without a memory, when correlations between system and environment exist only on time scales which are much shorter than the time scale on which the system evolves. It has therefore been argued that it is not possible to apply the Born approximation without using in addition certain Markovian approximations \cite{Breuer}. When applying both approximations to atom-field systems, we obtain the standard Born-Markov master equation \cite{Hu2}. All possible representations of $H$ result in the same master equation, independent of whether or not the RWA is applied to the Hamiltonian. In other words, the only way of obtaining a master equation which is more precise than already existing master equations is to allow for non-Markovian descriptions (cf.~eg.~Refs.~\cite{Petro2,Maniscalco,Hu}). However, although these papers contain interesting results for a variety of quantum systems, they do not apply to the system which we consider here.

The reason for this is that the above observation only applies when system and bath form a closed quantum system with only unitary dynamics, as it is widely assumed in the literature \cite{noise,Mandel,Cohen2,Walls,Caldeira,Breuer}. Here we adopt a different point of view. Following the ideas of Hegerfeldt and Wilser \cite{Wilser,Hegerfeldt93} and others \cite{Molmer,Carmichael,Zoller}, we take the interaction of the free radiation field with an external environment, like a detector or the walls of the laboratory, explicitly into account. Following the ideas of Zurek and Paz on decoherence and einselection (see eg.~Ref.~\cite{Zurek2}) and applying them to the free radiation field and its external environment, we assume that this environment resets the free radiation field on a coarse grained time scale $\Delta t$ onto its vacuum state $|0 \rangle$. This means, the Born approximation which assumes that the density matrix of the atom-field system can be written as
\begin{eqnarray} \label{rhonew}
\rho (t) &=& \rho_{\rm A} (t) \otimes \ket{0}\bra{0} \, ,
\end{eqnarray}
is no longer an approximation but the result of external environmental couplings. Hence it is well justified to derive an effective master equation for the atomic system without employing the RWA and certain Markovian approximations, but still employing Eq.~(\ref{rhonew}). As a result, we obtain a master equation whose constants depend on the typical environmental response time $\Delta t$ and which is automatically Markovian. This is as expected, since the free radiation field does not constitute a memory for the states of the atomic system.

The above described approach of deriving master equations for atomic systems with spontaneous photon emission is in very good agreement with actual quantum optics experiments \cite{Zeno,Zeno2,Zeno3,Eichmann}. For example, it has been shown experimentally that certain three-level atoms exhibit macroscopic quantum jumps \cite{Zeno,Zeno2,Zeno3}, ie.~they emit a completely random telegraph signal with long intervals of intense photon emissions interrupted by periods of the complete absence of photons \cite{Beige,Berkeland}. In the experiment by Eichmann {\em et al.} \cite{Eichmann} two laser driven two-level atoms emit photons onto a far-away screen, where they form an interference patterns which is typical for double slit experiments \cite{Schon}. In both experiments, the free radiation field constantly leaks information about the atomic system into an external environment. When an atom emits a real photon, the photon flies away and is absorbed after a relatively short time $\Delta t$. It does not return to interact again with its source. 

As summarised in Eq.~(\ref{rhonew}), environmental couplings project the atom-field system onto a product state with the free radiation field in its vacuum state. Notice however that for most representations of the atom-field system, the ground state of the Hamiltonian $H$ in Eq.~(\ref{h1}) is an entangled state. This means, in most derivations of master equations, the interaction with the environment resets the atom-field system constantly onto a state which is energetically higher than its ground state. If this applies, the photon-absorbing environment constantly pumps energy into the system which then manifests itself as a non-zero stationary state photon emission rate even in the absence of external driving \cite{Schulman,Andreas2}. Such a concentration of energy might play a crucial role in sonoluminescence experiments \cite{SL}. However, it need not play a role in atom-field systems, since the ground state of the rotating wave Hamiltonian is a product state with the free radiation field in its vacuum state.

Some authors argue that the description of a physical system like an atom in a free radiation field should be gauge-independent (see eg.~Ref.~\cite{Chiao} and references therein). This is no longer the case when we select for example the master equation associated with the rotating wave Hamiltonian. However, notice that we consider the atom-field system as an open quantum system with a photon-absorbing environment. This environment makes a concrete choice as to which component of the atom-field system represents the ``field" and which one represents the "atom." A similar choice should be made when deriving an atomic master equation.

There are five sections in this paper. In Section \ref{section2} we introduce a large class of unitarily equivalent representations of the atom-field Hamiltonian. In Section \ref{section3}, we obtain a large set of non-equivalent master equations. In Section \ref{section4}, we calculate the corresponding stationary state photon emission rate. The results of this calculation help us to select what we believe to be the most accurate master equation. Finally, we summarise our findings in Section \ref{conc}. 

\section{Unitarily equivalent Hamiltonians} \label{section2}

In this section we focus our attention on a large set of unitarily equivalent atom-field Hamiltonians which we obtain by applying a wide range of unitary transformations $R$ to the {\em minimal coupling} Hamiltonian. For example, the Power-Zienau-Woolley (PZW) transformation transforms this Hamiltonian into the so-called {\em multipolar} Hamiltonian \cite{cohen1,woolley1,woolley2}. In general, we obtain a Hamiltonian which is a mixture of the minimal coupling and the multipolar Hamiltonians \cite{baxter}. The most notable of these Hamiltonians is the one that does not contain any counter rotating terms in the atom-field interaction \cite{drummond}. Since this Hamiltonian is of the same form as the Hamiltonian $H$ of the atom-field system obtained when applying the RWA, we refer to it in the following as the {\em rotating wave} Hamiltonian. This Hamiltonian is now no longer the result of approximations but the result of a certain unitary transformation $R$.  

\subsection{The minimal coupling Hamiltonian}

To begin with we consider an electron of charge $e$ and mass $m$ which is trapped inside a potential $V({\bf r})$ and that couples to a continuum of quantised electromagnetic field modes. The canonical operators of the radiation field are the transverse vector potential ${\bf A}$ and its conjugate momentum ${\bf \Pi}$. These satisfy the transverse canonical commutation relation 
\begin{eqnarray} \label{conjugate}
[{\rm A}_i({\bf x}),\Pi_j({\bf x}')] &=& {\rm i} \hbar \, \delta_{\it ij}^{\rm T}({\bf{x}}-{\bf{x}}') \, , 
\end{eqnarray}
where the indices $i$ and $j$ specify different spatial components. The canonical operators for the electron are its position ${\bf{r}}$ and its conjugate momentum ${\bf{p}}$ with the canonical commutation relation
\begin{eqnarray}
[r_i,p_j] &=& {\rm i} \hbar\delta_{\it ij} \, . 
\end{eqnarray}
Using this notation, the minimal coupling Hamiltonian $H_{\rm min}$ can be written as
\begin{eqnarray} \label{min}
H_{\rm min} &=& {1 \over 2m} \left[ {\bf p} + e {\bf A} ({\bf 0}) \right]^2 + V({\bf r}) \nonumber \\ 
&& + \frac{1}{2} \int {\rm d}^3 {\bf{x}} \, \left[ {1 \over \mu_0} \, |{\bf B}({\bf x})|^2 + {1 \over \epsilon_0} |{\bf \Pi}({\bf x})|^2 \right] \, .
\end{eqnarray}
The integration in this equation is over the whole space occupied by the atom-field system and ${\bf B} = \nabla \times {\bf A}$ denotes the magnetic field. The first term on the right hand side of Eq.~(\ref{min}) has already been simplified with the help of the electric dipole approximation which replaces the position operator ${\bf r}$ of the electron in the vector potential ${\bf A}$ with the position ${\bf 0}$ of the centre of mass of the particle.

One can easily check that two operators ${\bf A}$ and ${\bf \Pi}$ defined as
\begin{eqnarray} \label{add}
{\bf A}({\bf x}) &=& \sum_{{\bf k}\lambda } \left({ \frac{\hbar}{2\epsilon_0\omega_kL^3}}\right)^{1 \over 2}  \, {\rm{\bf e}}_{{\bf k}\lambda } \, a_{{\bf k}\lambda } \, {\rm e}^{{\rm i} {\bf k} \cdot {\bf x}}+ {\rm H.c.} \, , \notag \\
{\bf \Pi}({\bf x}) &=& - {\rm i} \sum_{{\bf k}\lambda } \left({ \frac{\hbar \epsilon_0 \omega_k }{2 L^3}}\right)^{1 \over 2} {\rm {\bf e}}_{{\bf k}\lambda } \, a_{{\bf k}\lambda } \, {\rm e}^{{\rm i}{\bf k}\cdot{\bf x}} + {\rm H.c.}
\end{eqnarray}
obey the canonical commutator relation in Eq.~(\ref{conjugate}), if the $a_{{\bf k}\lambda}$ and $a_{{\bf k}\lambda }^\dagger$ are creation and annihilation operators with the bosonic commutation relation
\begin{eqnarray}\label{craadagger}
[a_{{\bf k}\lambda },a_{{\bf k}'\lambda '}^\dagger] &=& \delta_{{\bf k}{\bf k}'} \, \delta_{\lambda\lambda'} 
\end{eqnarray}
and with an appropriate choice of boundary conditions. Here $L^3$ denotes the quantisation volume of the free radiation field. Each field mode $({\bf k},\lambda)$ is characterised by a polarisation $\lambda$, a wavevector ${\bf k}$, and orthogonal unit polarisation vectors ${\rm{\bf e}}_{{\bf k}\lambda }$ with ${\rm{\bf e}}_{{\bf k} 1}$, ${\rm{\bf e}}_{{\bf k} 2}$, and ${\bf k}$ all being pairwise orthogonal. The corresponding magnetic field ${\bf B}$ is in addition given by
\begin{eqnarray} \label{addx}
{\bf B}({\bf x}) &=& {\rm i} \sum_{{\bf k}\lambda } \left({ \frac{\hbar}{2\epsilon_0\omega_kL^3}}\right)^{1 \over 2}  \, ( \hat {\bf k} \times {\rm{\bf e}}_{{\bf k}\lambda } ) \, a_{{\bf k}\lambda } \, {\rm e}^{{\rm i} {\bf k} \cdot {\bf x}} \nonumber \\
&& + {\rm H.c.} \, ,
\end{eqnarray}
where $\hat {\bf k} = {\bf k}/ k$ is a normalised vector. Eqs.~(\ref{add})--(\ref{addx}) can now be used to bring the minimal coupling Hamiltonian into a more compact form, 
\begin{eqnarray} \label{min3}
H_{\rm min} &=& {1 \over 2m} \left[ {\bf p} + e {\bf A} ({\bf 0}) \right]^2 + V({\bf r}) + \sum_{{\bf k}\lambda } \hbar \omega_k \, a_{{\bf k}\lambda }^\dagger a_{{\bf k}\lambda }  \, , \nonumber \\
\end{eqnarray}
where only an overall constant with no physical consequences has been neglected.
 
\subsection{Alternative representations of the atom-field Hamiltonian}

As suggested by Drummond \cite{drummond}, we now consider the unitary transformation $R_{\{\alpha_k\}}$ in the electric dipole approximation which is defined as
\begin{eqnarray}\label{ralphak}
R_{\{\alpha_k\}} &\equiv& \exp \left({{\rm i} e \over \hbar} {\bf A}_{\{\alpha_k\}}({\bf 0}) \cdot {\bf r} \right) \, .
\end{eqnarray}
The field operator ${\bf A}_{\{\alpha_k\}}({\bf 0})$ in this equation is given by
\begin{eqnarray} \label{add2}
{\bf A}_{\{\alpha_k\}}({\bf 0}) = \sum_{{\bf k}\lambda } \left({ \frac{\hbar}{2\epsilon_0\omega_kL^3}}\right)^{1 \over 2}  \, \alpha_k \, {\rm{\bf e}}_{{\bf k}\lambda } \, a_{{\bf k}\lambda } + {\rm H.c.} 
\end{eqnarray}
with the $\alpha_k$'s being real and dimensionless. In the case, where all the $\alpha_k$ in Eq.~(\ref{ralphak}) equal one, the above transformation $R_{\{\alpha_k\}} $ reduces to the well-known PZW transformation in the electric dipole approximation \cite{cohen1,baxter,drummond}. 

We now apply $R_{\{ \alpha_k \} }$ in Eq.~(\ref{ralphak}) to the minimal coupling Hamiltonian in Eq.~(\ref{min3}). The details of this calculation which yields an $\{ \alpha_k \}$-dependent Hamiltonian of the same form as Eq.~(\ref{h1}) but with its components given by \cite{baxter}
\begin{eqnarray}\label{ham}
H_{\rm A} &=& {{\bf p}^2 \over 2m} +V({\bf r}) + \sum_{{\bf k}} {e^2 \over 2\epsilon_0 L^3} \, \alpha_k^2 \, |{\bf r} |^2 \, , \nonumber \\ 
H_{\rm F} &=& {e^2 \over 2m} \, |{\bf A}({\bf 0}) - {\bf A}_{\{ \alpha_k \} }({\bf 0})|^2 + \sum_{{\bf k}\lambda } \hbar \omega_k \, a_{{\bf k}\lambda }^\dagger a_{{\bf k}\lambda } \, , ~~~ \nonumber \\
H_{\rm AF} &=& {e \over m} \, {\bf p} \cdot \left( {\bf A}({\bf 0}) - {\bf A}_{\{ \alpha_k \} }({\bf 0}) \right) - {e\over \epsilon_0} \, {\bf \Pi}_{\{\alpha_k\}} ({\bf 0}) \cdot {\bf r} ~~ \nonumber \\
\end{eqnarray}
can be found in App.~\ref{app3}. The field operator ${\bf \Pi}_{\{\alpha_k\}}({\bf 0}) $ in this equation is given by 
\begin{eqnarray} \label{add2}
{\bf \Pi}_{\{\alpha_k\}}({\bf 0}) = - {\rm i} \sum_{{\bf k}\lambda } \left({ \frac{\hbar \epsilon_0 \omega_k}{2 L^3}}\right)^{1 \over 2}  \, \alpha_k \, {\rm{\bf e}}_{{\bf k}\lambda } \, a_{{\bf k}\lambda } + {\rm H.c.} ~~
\end{eqnarray}
The transformation $R_{\{ \alpha_k \}}$ does not leave the different components of the Hilbert spaces ${\mathcal H}_{\rm A}$ and ${\mathcal H}_{\rm F}$ of atom and field invariant. This is because the canonical momenta ${\bf p}$ and ${\bf \Pi}$ take on different physical meanings for each one of the possible representations of the atom-field Hamiltonian $H$.  Each representation refers to physically different components of the total system as representing the ``atom" and as representing the ``field." 

The bare Hamiltonians $H_{\rm A}$ and $H_{\rm F}$ in Eq.~(\ref{ham}) contain self-energy contributions. These are divergent without the introduction of an upper frequency cut-off for the modes of the free radiation field. In order to show that these divergences do not contribute to our derivation of master equations, we now consider an atomic two-level system with ground state $|1 \rangle$, excited state $|2 \rangle$, and an energy separation $\hbar \omega_0$. Moreover, we introduce the Pauli operators
\begin{eqnarray}
&& \sigma^+ = \ket{2}\bra{1} \, ,  ~~ \sigma^- = \ket{1}\bra{2} \, , \nonumber \\
&& \sigma_3 =  {1 \over 2} \left ( |2 \rangle \langle 2|-|1 \rangle \langle 1| \right )
\end{eqnarray}
with the spin-like commutation relations 
\begin{eqnarray} \label{SLCR}
[\sigma^+,\sigma^-] = 2 \sigma_3 \, , ~~ [\sigma^\pm,\sigma_3] = \mp \sigma^\pm \, . 
\end{eqnarray}
Using the commutation relation 
\begin{eqnarray}
[H_{\rm A},r_i] &=& - {{\rm i} \hbar \over m} \, p_i
\end{eqnarray}
and multiplying the atomic operators ${\bf r}$ and ${\bf p}$ on both sides with the identity $|1 \rangle \langle 1| + |2 \rangle \langle 2| = 1$, we find that the components $r_i$ and $p_i$ may be written as 
\begin{eqnarray}\label{mom}
r_i &=& d_i \, \sigma^+ + {\rm H.c.} \, ,\nonumber \\
p_i &=& {\rm i} m \omega_0 \, d_i \, \sigma^+ + {\rm H.c.} \, ,
\end{eqnarray}
where the $d_i$ are the components of the (in general complex) atomic dipole moment 
\begin{eqnarray} \label{bfd}
{\bf d} = \bra{2}{\bf r}\ket{1} \, . 
\end{eqnarray}
For a two-level atom, the atomic self-energy term merely shifts the zero-point energy, because it is diagonal in the bare atomic basis. Neglecting it is equivalent to neglecting a constant in the Hamiltonian. The field self-energy term does not contribute in the second order perturbation theory calculations in Section \ref{more}. It can therefore be neglected without affecting the results of this paper. Taking this into account, we obtain the general atom-field Hamiltonian 
\begin{eqnarray}\label{ham3}
H &=& \sum_{{\bf k}\lambda } \hbar g_{{\bf k} \lambda} \, \sigma^+ \left( u_k^+ \, a_{{\bf k}\lambda }^\dagger + u_k^- \, a_{{\bf k}\lambda } \right) + {\rm H.c.} \nonumber \\ 
&& + \hbar  \omega_0 \, \sigma_3 + \sum_{{\bf k}\lambda } \hbar  \omega_k \, a_{{\bf k}\lambda }^\dagger a_{{\bf k}\lambda } 
\end{eqnarray}
with the atom-field coupling constant $g_{{\bf k} \lambda}$ and the (real) coefficients $u_k^\pm$ defined as
\begin{eqnarray} \label{u-2}
g_{{\bf k} \lambda} &\equiv & {\rm i} e \left({\omega_0 \over 2\epsilon_0 \hbar L^3}\right)^{1\over 2} \, {\bf  e}_{{\bf k}\lambda }\cdot{\bf d}  \, , \nonumber \\ 
u_k^\pm &\equiv & (1-\alpha_k) \left( {\omega_0 \over \omega_k} \right)^{1/2} \mp \alpha_k \left( {\omega_k \over \omega_0} \right)^{1/2} \, .
\end{eqnarray}
When setting all $\alpha_k$ equal to zero, the minimal coupling Hamiltonian in the electric dipole approximation reduces to the minimal coupling form 
\begin{eqnarray} \label{min2}
H_{\rm min} &=& \sum_{{\bf k}\lambda } \hbar g_{{\bf k} \lambda} \left( {\omega_0 \over \omega_k} \right)^{1/2} \left( \sigma^+ + \sigma^- \right) \left( a_{{\bf k}\lambda } - a_{{\bf k}\lambda }^\dagger \right) \nonumber \\ 
&& + \hbar  \omega_0 \, \sigma_3 + \sum_{{\bf k}\lambda } \hbar  \omega_k \, a_{{\bf k}\lambda }^\dagger a_{{\bf k}\lambda }  
\end{eqnarray}
which contains a linear coupling between its respective atomic system and the modes $({\bf k},\lambda)$ of its respective free radiation field. Let us now have a closer look at two other examples of the general Hamiltonian.

\subsection{The multipolar Hamiltonian}

Another Hamiltonian which is often used in the literature is the multipolar Hamiltonian. It arises from the minimal coupling Hamiltonian after applying the PZW transformation. This means, the multipolar Hamiltonian for an atomic two-level system and in the electric dipole approximation corresponds to the Hamiltonian $H$ in Eq.~(\ref{ham3}) with all $\alpha_k$ identical to one. It equals
\begin{eqnarray} \label{mult}
H_{\rm mult} &=& \sum_{{\bf k}\lambda } \hbar g_{{\bf k} \lambda} \left( {\omega_k \over \omega_0} \right)^{1/2} \left( \sigma^+ + \sigma^- \right) \left( a_{{\bf k}\lambda } - a_{{\bf k}\lambda }^\dagger \right) \nonumber \\ 
&& + \hbar  \omega_0 \, \sigma_3 + \sum_{{\bf k}\lambda } \hbar  \omega_k \, a_{{\bf k}\lambda }^\dagger a_{{\bf k}\lambda } \, . 
\end{eqnarray}
Like the minimal coupling Hamiltonian, this representation implies a linear coupling between its respective atomic system and the modes $({\bf k},\lambda)$ of its respective free radiation field.

\subsection{The rotating wave Hamiltonian} \label{RWH}

The physical motivation for introducing the unitary transformation $R_{\{ \alpha_k \}}$ in Eq.~(\ref{ralphak}) is that this transformation allows us to remove the counter rotating terms of the form $\sigma^- a_{{\bf k}\lambda }$ and $\sigma^+ a_{{\bf k}\lambda }^\dagger$ from the atom-field Hamiltonian in electric dipole approximation \cite{drummond}. Indeed, when choosing the coefficients $\alpha_k$ such that
\begin{eqnarray} \label{alphak}
\alpha_k &=& {\omega_0 \over \omega_0 + \omega_k} \, ,
\end{eqnarray}
the general Hamiltonian $H$ in Eq.~(\ref{ham3}) simplifies to 
\begin{eqnarray}\label{ham4}
H_{\rm rot} &=& \sum_{{\bf k}\lambda } \hbar g_{{\bf k} \lambda} \,  {2 (\omega_0 \omega_k)^{1/2} \over \omega_0 + \omega_k} \, \sigma^+ a_{{\bf k}\lambda } + {\rm H.c.} \nonumber \\ 
&& + \hbar  \omega_0 \, \sigma_3 + \sum_{{\bf k}\lambda } \hbar  \omega_k \, a_{{\bf k}\lambda }^\dagger a_{{\bf k}\lambda } \, .
\end{eqnarray}
In the following, we refer to this Hamiltonian as the {\em rotating wave} Hamiltonian, since almost the same Hamiltonian is obtained when applying the RWA. The only difference between these Hamiltonians is a difference in the atom-field coupling constants. 

\section{Non-equivalent master equations}\label{section3}

In this section we derive the the master equation for the effective time evolution of an atomic system with spontaneous photon emission without applying the Born and certain Markovian approximations. Our only approximations are the use of second order perturbation theory within an appropriately chosen interaction picture and the resetting of the free radiation field on a coarse grained time scale $\Delta t$ into its vacuum state. The first assumption is well justified as long as the environmental response time $\Delta t$ is short compared to the characteristic time scale of the effective atomic evolution. The second approximation is justified in the presence of a photon-absorbing environment which monitors the free radiation field \cite{Hegerfeldt93,Molmer,Carmichael,Zoller}. Independent of whether or not a photon has been found, decoherence destroys any coherences between the radiation field and its surroundings and transfers it into an environmentally preferred (pointer or einselected) state -- the vacuum state $|0 \rangle$ \cite{Zurek2}.

\subsection{Interaction picture}

The starting point of our derivation is, as usual, the Hamiltonian $H$ in Eq.~(\ref{h1}). In this subsection, we transfer this Hamiltonian into the interaction picture with respect to the free Hamiltonian
\begin{eqnarray} \label{H0}
H_0 &=& \hbar  \omega_0 \, \sigma_3 + \sum_{{\bf k}\lambda } \hbar  \omega_k \, a_{{\bf k}\lambda }^\dagger a_{{\bf k}\lambda } \, .
\end{eqnarray}
Using Eq.~(\ref{ham3}), we find that the general atom-field interaction Hamiltonian $H_{\rm I}$ equals
\begin{eqnarray} \label{HI}
H_{\rm I} &=& \sum_{{\bf k}\lambda} \hbar g_{{\bf k} \lambda} \, \sigma^+ \, {\rm e}^{{\rm i}  \omega_0 t} \left( u_k^+ \, a_{{\bf k}\lambda }^\dagger \, {\rm e}^{{\rm i}  \omega_kt} + u_k^- \,  a_{{\bf k}\lambda } \, {\rm e}^{- {\rm i}  \omega_kt} \right) \nonumber \\ 
&& + {\rm H.c.} 
\end{eqnarray}
Within this interaction picture, the atomic density matrix $\rho_{\rm A}(t)$ in the Schr\"odinger picture becomes
\begin{eqnarray} \label{rho01xxx}
\rho_{\rm A I} (t) &=& U_0^\dagger (t,0) \, \rho_{\rm A} (t) \, U_0 (t,0) \, . 
\end{eqnarray} 
Defining the superoperator ${\cal L}$ such that
\begin{eqnarray} \label{grad}
\dot \rho_{\rm A}(t) &=& {\cal L} (\rho_{\rm A} (t)) 
\end{eqnarray} 
and taking the time derivative of Eq.~(\ref{rho01xxx}), we find that  
\begin{eqnarray} \label{rhodotI}
\dot \rho_{\rm A I}(t) &=& {\rm i}  \omega_0 \, [\sigma_3, \rho_{\rm A I}] \nonumber \\
&& \hspace*{-1cm} + U_0^\dagger (t,0) \, {\cal L} \left( U_0 (t,0) \rho_{\rm A I} U_0^\dagger (t,0) \right) \, U_0 (t,0) \, . ~~~
\end{eqnarray} 
In the following subsections, we calculate this time derivative using second order perturbation theory and then use Eq.~(\ref{rhodotI}) to find the superoperator ${\cal L}$ in the Schr\"odinger picture. 

\subsection{Photon-absorbing environment} \label{discuss}

Following ideas of Hegerfeldt and Wilser \cite{Wilser,Hegerfeldt93} and others \cite{Molmer,Carmichael}, we do not assume that the atomic system and the surrounding free radiation field are a closed quantum system with purely unitary dynamics. Instead, we take the coupling of the free radiation field to an additional external environment, like a detector or the walls of the laboratory, explicitly into account. This external environment thermalises very rapidly and is in general in an equilibrium state. We denote this equilibrium state in the following by $\rho^{\rm env}_{\rm ss}$. 

Suppose, the interaction between the atomic system and the free radiation field had already created a small photon population in the field modes $({\bf k}, \lambda)$. If the photons in the free radiation field are {\em real} photons, they fly away and reach the external environment, ie.~a detector or the walls of the laboratory, after a relatively short time. There they interact with a large collection of atoms such that
\begin{eqnarray} \label{60}
\rho_{\rm F} \otimes \rho^{\rm env}_{\rm ss} &\longrightarrow & |0 \rangle \langle 0| \otimes \rho^{\rm env} (\rho_{\rm F} ) \, , 
\end{eqnarray}
where $\rho_{\rm F}$ is the initial density matrix of the free radiation field and $|0 \rangle$ is its vacuum state with 
$a_{{\bf k} \lambda} \, |0 \rangle = 0$ for all ${\bf k}$ and $\lambda$. In other words, any initial excitation in the free radiation field vanishes very quickly as a result of the interaction between the radiation field and its external environment. Eq.~(\ref{60}) also assumes that there are no coherences between the free radiation field and its environment. These have already been destroyed by decoherence \cite{Zurek2}.

Moreover, the environment itself thermalises in general very rapidly. Taking this into account, the effect of the photon-absorbing environment can be summarised as
\begin{eqnarray} \label{61}
\rho_{\rm F} \otimes \rho^{\rm env}_{\rm ss} &\longrightarrow & |0 \rangle \langle 0| \otimes \rho^{\rm env}_{\rm ss} \, ,
\end{eqnarray}
if the free radiation field does not become re-populated in the process. This applies when the vacuum state $|0 \rangle$ is an einselected state of the free radiation field. Einselected (ie.~environmentally-selected) states are states which remain stable in spite of the environment \cite{Zurek2}. Here we assume that the free radiation field possesses only a single einselected state. This is justified on the grounds that emitted photons fly away from the atom and are absorbed. In the absence of an atomic emitter but in the presence of a photon-absorbing environment, the vacuum state $|0 \rangle$ is the only state of the free radiation field which does not evolve in time. 

\subsection{Derivation of master equations}

The above considerations allow us to write the density matrix $\rho (t)$ of the atom-field system at any time $t$ as 
\begin{eqnarray} \label{rho}
\rho_{\rm I} (t) &=& \rho_{\rm A I} (t) \otimes \ket{0}\bra{0} \, ,
\end{eqnarray}
where $\rho_{\rm A I} (t)$ denotes an atomic density matrix with respect to the above introduced interaction picture. Eq.~(\ref{rho}) is known as the Born approximation but, as we have seen above, the assumption of an uncorrelated atom-field state becomes exact in the presence of a photon-absorbing environment. 

The second step in our derivation of a master equation for the atomic system is the calculation of the time evolution of the state in Eq.~(\ref{rho}) over a time interval $\Delta t$. Within this time, $\rho_{\rm I} (t)$ in Eq.~(\ref{rho}) evolves into 
\begin{eqnarray} \label{rhoDeltat}
&& \hspace*{-1cm} \rho_{\rm I} (t+\Delta t) \nonumber \\
&=& U_{\rm I}(t+\Delta t,t) \, \rho_{\rm A I} (t) \otimes \ket{0}\bra{0} \, U_{\rm I}^\dagger (t+\Delta t,t) \, . ~~
\end{eqnarray}
This atom-field density matrix corresponds in general to an entangled state with population in most modes $({\bf k},\lambda)$ of the free radiation field. Combining Eqs.~(\ref{60}), (\ref{61}), and (\ref{rhoDeltat}), we find that the density matrix of atom, field, {\em and} environment evolves during any subsequent photon absorption according to
\begin{eqnarray} \label{rhoyyy}
&& \hspace*{-0.9cm} \rho_{\rm I} (t + \Delta t) \otimes \rho^{\rm env}_{\rm ss} \nonumber \\
&\longrightarrow & {\rm Tr}_{\rm F} \left[ U_{\rm I}(t+\Delta t,t) \, \rho_{\rm A I} (t) \otimes \ket{0}\bra{0} \, U_{\rm I}^\dagger (t+\Delta t,t) \right] \nonumber \\
&& \hspace*{0.5cm} \otimes |0 \rangle \langle 0| \otimes \rho^{\rm env}_{\rm ss} \, , ~~~~
\end{eqnarray}
where ${\rm Tr}_{\rm F}$ denotes the trace over all the modes of the free radiation field. This equation shows that the density matrix of atom and field remains uncorrelated on the coarse grained time scale given by $\Delta t$.

An alternative way of deriving Eq.~(\ref{rhoyyy}) is to assume an environment which performs rapidly repeated, photon-absorbing measurements on a coarse grained time scale $\Delta t$ \cite{Wilser,Hegerfeldt93}. Calculations based on the RWA \cite{Schon}, which reproduce the observations of recent quantum optics experiments like the one by Eichmann {\em et al.}~\cite{Eichmann}, show that it is sufficient to assume an environment which resolves the direction of each emitted photon but not their frequency in order to obtain Eq.~(\ref{rhoyyy}). Our derivation of master equations is hence consistent with the work by Hegerfeldt and Wilser \cite{Wilser,Hegerfeldt93} who derived master equations while assuming photon-absorbing broadband measurements. Like einselection, these environment-induced measurements destroy any correlations between the atomic system and the free radiation field. The trace over the free radiation field in Eq.~(\ref{rhoyyy}) assures that the density matrix on the right hand side is always normalised. Eq.~(\ref{rhoyyy}) also takes into account that the interaction between radiation field and its external environment occurs on a very short time scale without non-local effects on the atomic state.

Here we are only interested in the time evolution of $\rho_{\rm A I} (t)$. Eq.~(\ref{rhoyyy}) shows that this density matrix evolves such that 
\begin{eqnarray} \label{rhoADeltat2}
\rho_{\rm A I} (t+\Delta t) &=& \rho_{\rm A I}^0 (t+\Delta t) + \rho_{\rm A I}^> (t+\Delta t) \, ,  
\end{eqnarray} 
if we define
\begin{eqnarray} \label{rho01}
\rho_{\rm A I}^0 (t+\Delta t) &\equiv & \bra{0} \, U_{\rm I}(t+\Delta t,t) \, \ket{0} \, \rho_{\rm A I} (t) \nonumber \\
&& \hspace*{1.2cm} \times \bra{0} \, U_{\rm I}^\dagger (t+\Delta t,t) \, \ket{0} \, , \nonumber \\
\rho_{\rm A I}^> (t+\Delta t) &\equiv & \sum_{n=1}^\infty \sum_{{\bf k} \lambda} \bra{n_{{\bf k} \lambda}} \, U_{\rm I}(t+\Delta t,t) \, \ket{0} \, \rho_{\rm A I} (t) \nonumber \\
&& \hspace*{1.2cm} \times \bra{0} \, U_{\rm I}^\dagger (t+\Delta t,t) \, \ket{n_{{\bf k} \lambda}} \, . ~~~~
\end{eqnarray} 
These atomic density matrices, respectively, describe the subensemble of atoms {\em without} and the subensemble of atoms {\em with} photon emission in $(t,t+\Delta t)$. 

The purpose of an atomic master equation is to establish a direct connection between $\rho_{\rm A I} (t+\Delta t)$ in Eq.~(\ref{rhoADeltat2}) and $\rho_{\rm A I} (t)$ in Eq.~(\ref{rho}) without having to calculate the trace over the free radiation field after each time step $\Delta t$. To create such a link, we calculate the difference quotient
\begin{eqnarray} \label{grad2}
{\cal L}_{\rm I} (\rho_{\rm A I} (t)) &\equiv & {\rho_{\rm A I} (t+\Delta t) -\rho_{\rm A I} (t) \over \Delta t } \, .
\end{eqnarray}
In the limit $\Delta t \to 0$, this expression becomes a time derivative, giving the relation
\begin{eqnarray} \label{zzz}
\dot \rho_{\rm A I}(t) &=& {\cal L}_{\rm I} (\rho_{\rm A I} (t)) \, .
\end{eqnarray} 
However, since the superoperator ${\cal L}_{\rm I}$ is obtained via a partial trace, by applying this limit one enters a regime in which the dynamics of the atom-field system becomes restricted onto the zero photon subspace \cite{Hegerfeldt93}. We therefore avoid this limit and only assume that the time increment $\Delta t$ is sufficiently small compared to the effective time scale over which the atomic system evolves. Eq.~(\ref{grad2}) now constitutes a Markovian approximation of the continuous dynamics given by Eq.~(\ref{zzz}).

\subsection{Second order perturbation theory} \label{more}

The next step in our derivation of an atomic master equation, ie.~of the superoperator ${\cal L}$ in Eq.~(\ref{grad}), is the explicit calculation of ${\cal L}_{\rm I}$ in Eq.~(\ref{zzz}). Since the time evolution of the atomic system is homogeneous in time, ${\cal L}_{\rm I}$ is independent of $t$ as long as no time-dependent interactions, like laser fields, are applied. However, one can easily check that ${\cal L}_{\rm I}$ depends explicitly on $\Delta t$. In the following, we treat $\Delta t$ as an external parameter which can be absorbed into the coefficients of the relevant master equation. 

\subsubsection{The subensemble without photon detection}

As mentioned already above, here we are especially interested in the case where $\Delta t$ is small compared to the time scale on which the atomic density matrix $\rho_{\rm A I}(t)$ evolves. This assumption allows us to calculate $\rho_{\rm A I}^0 (t+\Delta t)$ in Eq.~(\ref{rho01}) using second order perturbation theory. Doing so we find that
\begin{eqnarray}\label{uc1}
&& \hspace*{-1cm} \bra{0} U_{\rm I}(t + \Delta t,t) \ket{0} \nonumber \\
&=& 1 - {1 \over \hbar^2} \int_t^{t+\Delta{t}} {\rm d}t' \int_t^{t'} {\rm d}t'' \, \langle 0 | H_{\rm I} (t') H_{\rm I}(t'') |0 \rangle \, . ~~~
\end{eqnarray}
Using Eq.~(\ref{HI}), $\left(\sigma^\pm \right)^2 =  0$, and introducing the time independent constants $A'_{\pm}$ as
\begin{equation}\label{A'1}
A'_{\pm} \equiv {1 \over \Delta t} \int_0^{\Delta t} \! {\rm d}t' \int_0^{t'} \! {\rm d}t'' \, \sum_{{\bf k}\lambda}  |g_{{\bf k}\lambda}|^2 {u_k^{\pm}}^2  {\rm e}^{\mp {\rm i}( \omega_0 \pm  \omega_k)(t'-t'')} ,
\end{equation}
the above expression simplifies to 
\begin{eqnarray}\label{uc1yyy}
\hspace*{1cm} && \hspace*{-2cm} \bra{0} U_{\rm I}(t + \Delta t,t) \ket{0} \nonumber \\
&=& 1 - \left( A'_{-} \, \sigma^+\sigma^- + A'_{+} \, \sigma^-\sigma^+ \right) \Delta t \, . 
\end{eqnarray}
The constants $A'_{\pm}$ in Eq.~(\ref{A'1}) are in general complex numbers. Since their real and their imaginary parts have different physical meanings, it is convenient to consider them separately and to define
\begin{eqnarray}\label{A1}
A_{\pm} &\equiv & 2 \, {\rm Re} (A_\pm ')  \, , \nonumber \\ 
\Delta \omega_{1,2} &\equiv & {\rm Im} (A_\pm ') 
\end{eqnarray}
such that $A'_{\pm} = {1\over 2} A_{\pm} +{\rm i} \, \Delta \omega_{1,2}$. Substituting Eq.~(\ref{uc1}) into Eq.~(\ref{rho01}) and using this notation, we find that 
\begin{eqnarray}\label{rho02}
\rho_{\rm A I}^0(t+\Delta t) &=& \rho_{\rm A I}(t) + {\rm i} \sum_{i=1}^2 \Delta \omega_i \, [ \rho_{\rm A I}(t) , \ket{i} \bra{i} ] \, \Delta t \nonumber \\ 
&& \hspace*{-0.5cm} - {1 \over 2} \big(A_-\sigma^+\sigma^- + A_+\sigma^-\sigma^+\big) \rho_{\rm A I}(t) \, \Delta t  \nonumber \\
&& \hspace*{-0.5cm} - {1 \over 2} \rho_{\rm A I}(t) \, \big(A_-\sigma^+\sigma^- + A_+\sigma^-\sigma^+\big) \, \Delta t ~~~~~
\end{eqnarray}
up to first order in $\Delta t$. This means, the imaginary parts of $A'_+$ and $A'_-$ shift respectively the frequencies of the atomic states $|1 \rangle$ and $|2 \rangle$, while the real parts of these constants are decay rates. 

\subsubsection{The subensemble with photon detection} \label{r1}

Evaluating $U_{\rm I}(t+\Delta t,t)$ again using second order perturbation theory, the density matrix $\rho_{\rm A I}^>(t+\Delta t)$  in Eq.~(\ref{rho01}) for the subensemble with photon emission in $(t+\Delta t,t)$ becomes
\begin{eqnarray}\label{rho12}
\rho_{\rm AI}^>(t+\Delta t) &=& {1 \over \hbar^2} \sum_{{\bf k}\lambda} \int_t^{t+\Delta t} \! {\rm d}t' \int_t^{t + \Delta t} \! {\rm d}t'' \nonumber \\
&& \hspace*{-0.5cm} \bra{1_{{\bf k}\lambda}} H_{\rm I}(t') \ket{0} \, \rho_{\rm AI}(t) \, \bra{0} H_{\rm I}(t'') \ket{1_{{\bf k}\lambda}} \, . ~~
\end{eqnarray}
Substituting Eq.~(\ref{HI}) into this equation, $\rho_{\rm AI}^>(t+\Delta t)$ simplifies to 
\begin{eqnarray}\label{rho13}
\rho_{\rm AI}^>(t+\Delta t) &=& {\mathcal R}_{\rm I} (\rho_{\rm AI}(t)) \, \Delta t \, ,
\end{eqnarray}
having defined the reset superoperator ${\mathcal R}_{\rm I}$ as  
\begin{eqnarray}\label{rst}
{\mathcal R}_{\rm I} (\rho_{\rm A I}) &=& A_+ \, \sigma^+ \rho_{\rm A I} \sigma^- + A_- \, \sigma^- \rho_{\rm A I} \sigma^+  \nonumber \\ 
&& + B_+ \, \sigma^+ \rho_{\rm A I} \sigma^+ + B_- \, \sigma^- \rho_{\rm A I} \sigma^- 
\end{eqnarray}
with 
\begin{eqnarray}\label{b}
A_{\pm} &\equiv & \sum_{{\bf k}\lambda} |g_{{\bf k}\lambda}|^2 \, {u_k^{\pm}}^2 \, \Delta t \, {\rm sinc}^2 \left[ {1 \over 2} ( \omega_0 \pm  \omega_k) \Delta t \right] \, , ~~~
\end{eqnarray}
where ${\rm sinc} \, x \equiv \sin x/x$, and with
\begin{eqnarray}\label{b2}
B_+ &\equiv& {1 \over \Delta t} \int_t^{t+\Delta{t}} {\rm d}t' \int_t^{t+\Delta t} {\rm d}t'' \sum_{{\bf k}\lambda} g_{{\bf k}\lambda}^2 \, u_k^+ u_k^-  \nonumber \\
&& \hspace*{2cm}  \times {\rm e}^{{\rm i}  \omega_0(t'+t'')} \, {\rm e}^{{\rm i}  \omega_k (t'-t'')} \, , \nonumber \\
B_- &\equiv& {1 \over \Delta t} \int_t^{t+\Delta{t}} {\rm d}t' \int_t^{t+\Delta t} {\rm d}t'' \sum_{{\bf k}\lambda} \left(g_{{\bf k}\lambda}^*\right)^2 \, u_k^+ u_k^-  \nonumber \\
&& \hspace*{2cm}  \times {\rm e}^{-{\rm i}  \omega_0(t'+t'')} \, {\rm e}^{{\rm i}  \omega_k (t'-t'')} \, .
\end{eqnarray}
Using the specific properties of trigonometric functions, one can show that the $A_\pm$ in this equation are twice the real parts of the constants $A'_\pm$ in Eq.~(\ref{A'1}) and hence the same as the $A_\pm$ which we introduced in Section \ref{more}. 

\subsubsection{The superoperator ${\cal L}_{\rm I}$}

Substituting Eqs.~(\ref{rho02}), (\ref{rho13}), and (\ref{rst}) into Eq.~(\ref{grad2}), we find that the superoperator ${\cal L}_{\rm I}$ in Eq.~(\ref{grad2}) equals 
\begin{eqnarray} \label{rhodotI2}
{\cal L}_{\rm I} (\rho_{\rm A I}) &=& {\rm i} \sum_{i=1}^2 \Delta \omega_i \, [ \, \rho_{\rm A I} , \ket{i} \bra{i} \, ] \nonumber \\
&& + {1 \over 2} A_+ \left( 2 \sigma^+ \rho_{\rm A I} \sigma^- - \sigma^-\sigma^+ \rho_{\rm A I} - \rho_{\rm A I} \sigma^-\sigma^+ \right) \nonumber \\
&& +  {1 \over 2} A_- \left( 2 \sigma^- \rho_{\rm A I} \sigma^+ - \sigma^+\sigma^- \rho_{\rm A I} - \rho_{\rm A I} \sigma^+\sigma^- \right) \nonumber \\ 
&& + B_+ \, \sigma^+ \rho_{\rm A I} \sigma^+ + B_- \, \sigma^- \rho_{\rm A I} \sigma^- \, .
\end{eqnarray} 
The only approximation made in the derivation of this equation is the use of second order perturbation theory. 

\begin{figure}[t]
\begin{minipage}{\columnwidth}
\begin{center}
\hspace*{-1.5cm} \includegraphics[scale=1]{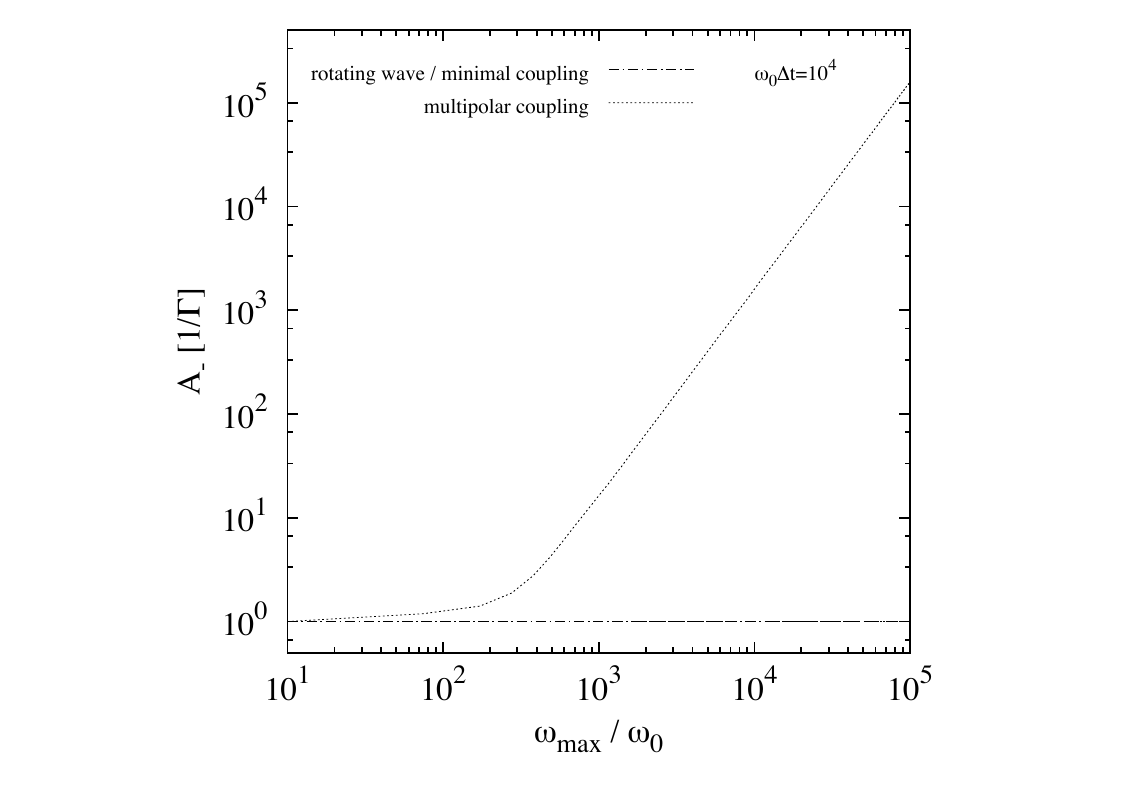}
\end{center}
\caption{Logarithmic plot of the transition rate $A_-$ in Eq.~(\ref{rates}) for the minimal coupling, the multipolar and the rotating wave Hamiltonian as a function of the upper cut-off frequency $\omega_{\rm max}$, while $\omega_{\rm min} = 0$ and $\omega_0 \Delta t = 10^4$.} \label{aminus}
\end{minipage}
\end{figure}

\begin{figure}[t]
\begin{minipage}{\columnwidth}
\begin{center}
\hspace*{-1.5cm} \includegraphics[scale=1]{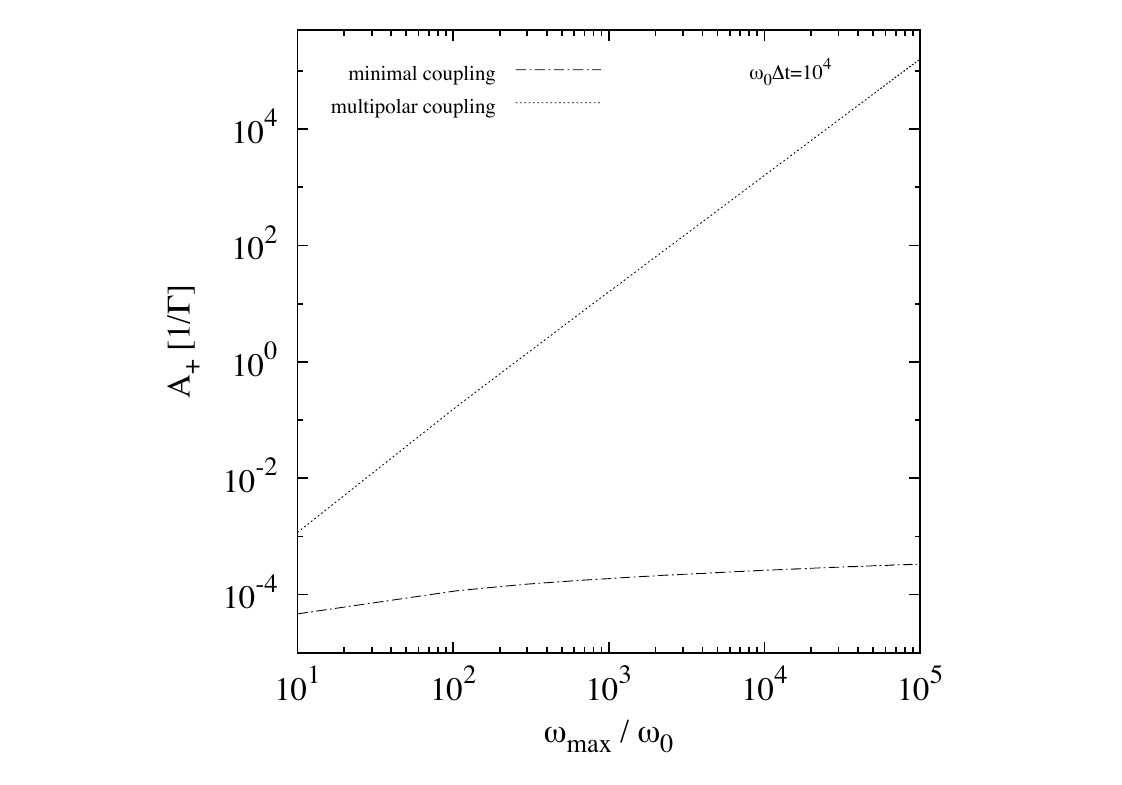}
\end{center}
\caption{Logarithmic plot of the transition rates $A_+$ in Eq.~(\ref{rates}) for the minimal coupling and the multipolar Hamiltonian as a function of the upper cut-off frequency $\omega_{\rm max}$, while $\omega_{\rm min} = 0$ and $\omega_0 \Delta t = 10^4$. These are in very good agreement with the analytical results in Eqs.~(\ref{Aminu})--(\ref{Amin2}).} \label{aplus}
\end{minipage}
\end{figure}

\subsection{General master equation}

To determine the superoperator ${\cal L}$ in the Schr\"odinger picture, we use Eq.~(\ref{rhodotI}) which implies
\begin{eqnarray} \label{rhodotI3}
{\cal L} (\rho_{\rm A}) &=& - {\rm i} \omega_0 \, [\sigma_3, \rho_{\rm A}] \nonumber \\
&& \hspace*{-1cm} + U_0 (t,0) {\cal L}_{\rm I} \left(U_0^\dagger (t,0) \rho_{\rm A} U_0 (t,0) \right) U_0^\dagger  (t,0) \, . ~~~
\end{eqnarray} 
Combining this equation with Eq.~(\ref{rhodotI2}) and redefining the atomic transition frequency $\omega_0$ such that the first term in Eq.~(\ref{rhodotI2}) can be absorbed into the free energy of the atom, we are finally able to calculate the superoperator ${\cal L}$. As a result we obtain the general master equation
\begin{eqnarray}\label{mstreq2}
\dot{\rho}_{\rm A} &=& - {\rm i} \tilde \omega_0 \, [ \sigma_3 , \rho_{\rm A} ] \nonumber \\ 
&& + {1 \over 2} A_+ \left( 2 \sigma^+ \rho_{\rm A} \sigma^- - \sigma^-\sigma^+ \rho_{\rm A} - \rho_{\rm A} \sigma^-\sigma^+ \right) \nonumber \\
&& +  {1 \over 2} A_- \left( 2 \sigma^- \rho_{\rm A} \sigma^+ - \sigma^+\sigma^- \rho_{\rm A} - \rho_{\rm A} \sigma^+\sigma^- \right) \nonumber \\ 
&& + B \, \sigma^+ \rho_{\rm A} \sigma^+ + B^* \, \sigma^- \rho_{\rm A} \sigma^- \, ,
\end{eqnarray}
where $\tilde \omega_0$ is a shifted atomic frequency and where $B$ is defined as 
\begin{eqnarray}\label{b2}
B &\equiv& \sum_{{\bf k}\lambda} g_{{\bf k}\lambda}^2 \, u_k^+ u_k^- \, \Delta t \, {\rm e}^{- {\rm i} \omega_0 \Delta t} \, {\rm sinc} \left[ {1 \over 2} (\omega_0 + \omega_k ) \Delta t \right] \nonumber \\
&& \times {\rm sinc} \left[ {1 \over 2} (\omega_0 - \omega_k ) \Delta t \right]  \, . ~~
\end{eqnarray}
Eq.~(\ref{mstreq2}) is the main result of this paper. Having a closer look at the definitions of $A_\pm$ and $B$, we see that $\dot \rho_{\rm A}$ is indeed time-independent, ie.~Markovian.

A relatively detailed analysis of the transition rates $A_-$ and $A_+$ can be found in App.~\ref{app}. However, performing the integration in Eq.~(\ref{rates}) is only possible after choosing concrete values for the constants $\alpha_k$ in Eq.~(\ref{u-2}). We therefore restrict ourselves in the following to the three most prominent Hamiltonians: the rotating wave, the minimal coupling, and the multipolar Hamiltonians. Figs.~\ref{aminus} and \ref{aplus} show $A_-$ and $A_+$ as a function of $\omega_{\rm max}$. Since $A_-$ and $A_+$ depend both only weakly on $\omega_{\rm min}$, we assume here that $\omega_{\rm min} = 0$. For the minimal coupling and the rotating wave Hamiltonians, $A_-$ is essentially the same as the usual spontaneous decay rate $\Gamma$ in Eq.~(\ref{Gamma}). However, for the multipolar Hamiltonian, $A_-$ is in general much larger than $\Gamma$, even for relatively small values of the upper cut-off frequency $\omega_{\rm max}$. Fig.~\ref{aplus} illustrates that the counter-rotating terms, like $\sigma^+ a_{{\bf k} \lambda}^\dagger $, in the general atom-field interaction Hamiltonian $H_{\rm I}$ in Eq.~(\ref{HI}) can have a huge effect. For the multipolar Hamiltonian, we obtain transition rates $A_+$ much larger than $\Gamma$. Only for the rotating wave Hamiltonian, do we have that $A_+ \equiv 0$ (cf.~Eq.~(\ref{Aminu})). 

App.~\ref{a1} shows that the master equations associated with the rotating wave, the minimal coupling, and the multipolar Hamiltonians can be written in Lindblad form \cite{Lindblad} for the experimental parameters which we consider in this paper. This is important, since being of Lindblad form guarantees that the density matrices obtained from these equations are valid physical objects \cite{Bill,Rob}. The spontaneous emission of a photon can be caused by the de-excitation of the atomic system. But, due to the presence of the counter-rotating terms in the system Hamiltonian $H_{\rm I}$ in Eq.~(\ref{ham3}), it can also be caused by the simultaneous excitation of the atom and the free radiation field. 
 
\section{Feasible experimental tests} \label{section4}

In this section, we calculate the stationary state photon emission rate $I_{\rm ss}$ in the absence of external driving for the minimal coupling, the multipolar, and the rotating wave Hamiltonians. As we shall see below, this rate can become enormously large, even when introducing physically meaningful environmental response times $\Delta t$ and appropriate cut-off frequencies $\omega_{\rm min}$ and $\omega_{\rm max}$ for the modes of the free radiation field. We then compare our predictions with the findings of actual experiments with single quantum dots and colour centers in diamond. 

\subsection{Detector-model dependent stationary state photon emission rates} \label{SSER}

Having a closer look at the derivation of a general master equation in Section \ref{section3}, we find that the probability density $I(t)$ for the emission of a photon at time $t$ equals the trace over the density matrix ${\mathcal R}_{\rm I} (\rho_{\rm A I})$ in Eq.~(\ref{rst}). Denoting the matrix elements of the density matrix $\rho_{\rm A}$ such that 
\begin{eqnarray} \label{ME}
\rho_{\rm A} &=& \left( \begin{array}{cc} \rho_{11} & \rho_{12} \\  \rho_{21} & \rho_{22} \end{array} \right) \, ,
\end{eqnarray}
we find that 
\begin{eqnarray} \label{I(t)}
I(t) &=& A_- \, \rho_{22}(t) + A_+ \, \rho_{11}(t) \, .
\end{eqnarray}
Using Eq.~(\ref{mstreq2}), we find that the matrix elements of $\rho_{\rm A}$ evolve according to 
\begin{eqnarray}
&& \dot \rho_{11} = - A_+ \, \rho_{11} + A_- \, \rho_{22} \, , \nonumber \\
&& \dot \rho_{12} = {\rm i}  \omega_0 \, \rho_{12} + {1 \over 2} (A_- + A_+) \, \rho_{12} + B^* \, \rho_{21} \, , \nonumber \\
&& \rho_{21} = \rho_{12}^* \, , ~~ \rho_{22} = 1 - \rho_{11} \, .
\end{eqnarray}
The stationary state of the atom ($\dot \rho_{\rm A} = 0$) corresponds to a diagonal matrix with
\begin{eqnarray}
\rho_{22} &=& {A_+ \over A_- + A_+} \, .
\end{eqnarray}
Substituting this population into Eq.~(\ref{I(t)}), we obtain the stationary state photon emission rate
\begin{eqnarray}\label{emrate2}
I_{\rm ss} &=& {2A_-A_+ \over  A_- + A_+} \, .
\end{eqnarray}
Half of these photons have a frequency such that they are {\em narrowband}. This means, their frequency is close to the atomic transition frequency $\omega_0$. The reason for this is that the stationary state photon emission rate from the excited atomic state $|2 \rangle$ equals $A_- \, \rho_{22}$, ie.~${1 \over 2} I_{\rm ss}$. This expression is proportional to $A_+$ (which is the rate associated with the broadband emission of photons), since the off-resonant excitation of the atomic system is what creates the stationary state population $\rho_{22} \neq 0$ in the first place.

\begin{figure}[t]
\begin{minipage}{\columnwidth}
\begin{center}
\hspace*{-1.5cm} \includegraphics[scale=1]{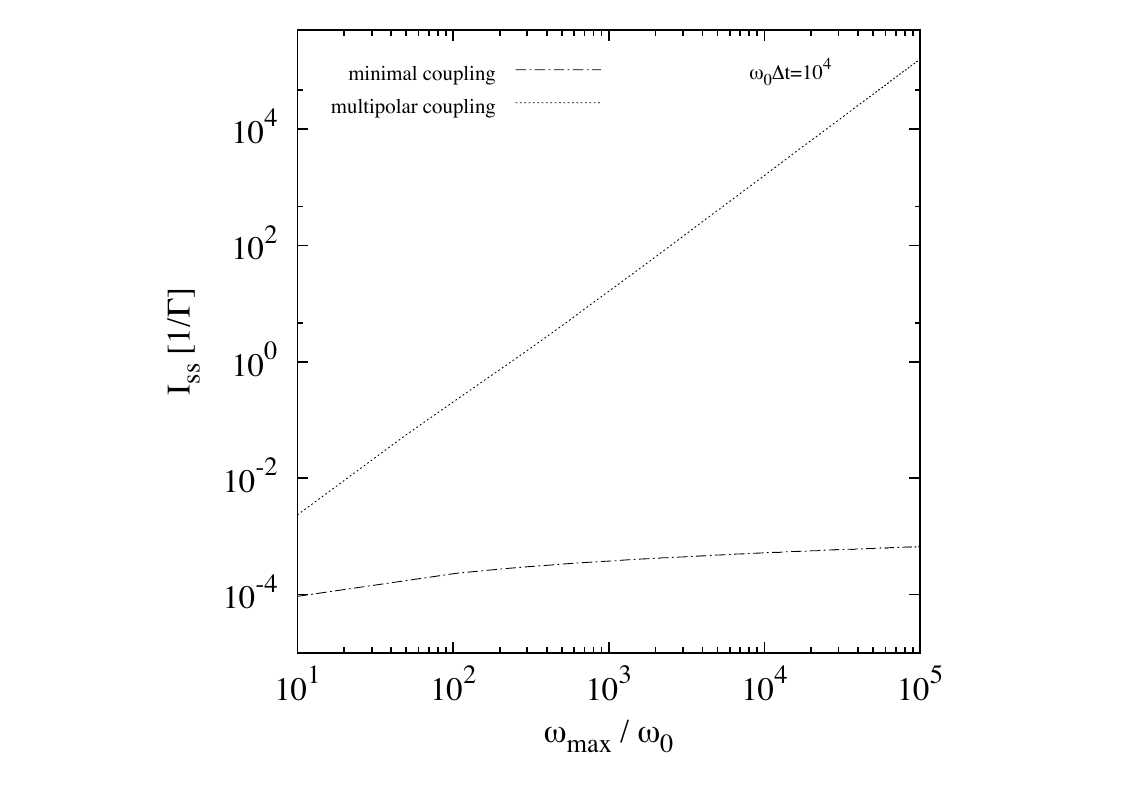}
\end{center}
\caption{Logarithmic plot of the stationary state photon emission rate $I_{\rm ss}$ for the minimal coupling and the multipolar Hamiltonian as a function of $\omega_{\rm max}$, while $\omega_{\rm min} = 0$. The graphs are the result of a numerical solution of Eqs.~(\ref{emrate2}) and (\ref{rates}).} \label{Iss}
\end{minipage}
\end{figure}

Fig.~\ref{Iss} shows the stationary state photon emission rate $I_{\rm ss}$ in Eq.~(\ref{emrate2}) for the minimal coupling and the multipolar Hamiltonians as a function of the cut-off frequency $\omega_{\rm max}$, while $\omega_{\rm min} = 0$. In case of the minimal coupling Hamiltonian we have $A_+ \ll A_-$ and $I_{\rm ss}$ is to a very good approximation given by $2 A_+$. In case of the multipolar Hamiltonian, $A_+$ and $A_-$ are approximately of the same size and $I_{\rm ss}$ equals $A_+$ for a wide range of cutoff frequencies. This is why Figs.~\ref{aplus} and \ref{Iss} seem to be essentially the same. However, the most surprising result in Fig.~\ref{Iss} is the actual size of $I_{\rm ss}$.

\subsection{The minimal coupling and the multipolar Hamiltonians} \label{figam}

We now have a closer look at realistic experimental parameters. A typical transition frequency in the optical regime is $\omega_0 = 3.7 \cdot 10^{15} \, $Hz. This frequency corresponds to the wavelength of $500 \,$nm. A possible estimate for the upper cut-off frequency for the modes of the free radiation field is $\omega_{\rm max} = 3.7 \cdot 10^{19} \, $Hz. This frequency corresponds to the Bohr radius \footnote{It is worth noticing that changing $\omega_{\rm max}$ by a few orders of magnitude does not change $A_+$ considerably.}. As we have seen above, a good estimate for the environmental-response time $\Delta t$ is the time it takes a photon to reach the walls of the laboratory. Assuming that photons travel at the speed of light and that the walls of the laboratory are about $3 \, {\rm m}$ away yields $\Delta t = 1 \cdot 10^{-8} \, {\rm s}$. Moreover, we assume in the following that $\Gamma = 1 \cdot 10^7 \, $Hz. Using these estimates and the equations for the minimal coupling Hamiltonian, we obtain a stationary state photon emission rate $I_{\rm ss}$ of about 1.5 photons per second. This rate is well below typical detector dark count rates of about 500 photons per second and more or less impossible to observe experimentally.

Substituting the same experimental parameters into the equations for the multipolar Hamiltonian, we obtain an $A_+$ of about $4 \cdot 10^6$ photons per second. This clearly shows that the master equation associated with the multipolar Hamiltonian cannot be valid. In other words, the multipolar Hamiltonian does not identify the free radiation field seen by a photon-absorbing environment correctly and corresponds instead to a highly unrealistic detector model. The photon emission rate predicted by the master equation associated with this Hamiltonian would have been noticed already in standard experiments with trapped ions. 

One way of obtaining higher stationary state photon emission rates even in case of the minimal coupling Hamiltonian is to move the photon-absorbing environment, ie.~the detector, closer to the atom. The reason for this is that $A_+$ scales essentially as $1/\Delta t$, while $A_-$ remains more or less the same when $\Delta t$ changes. For example, a photon-absorbing environment $10\,$cm away from the atom implies $\Delta t = 3 \cdot 10^{-10} \,$s and the above values yield an $I_{\rm ss}$ of about $40$ photons per second. Closer detectors yields larger $I_{\rm ss}$. We therefore now look for atomic systems with a relatively large spontaneous decay rate $\Gamma$ and a relatively small transition frequency $\omega_0$, like the ones used in recent experiments with single quantum dots and single colour centers in diamond. These are artificial atoms in the sense that each of them confines a single electron such that its states become quantised with level spacings in the optical regime.

For example, Matthiesen, Vamivakas, and Atat\"ure \cite{Atature} studied the fluorescence from Gallium Arsenide with a single layer of self-assembled Indium Arsenide at a temperature of $4.2 \,$K and measured a lifetime of $T_1 = 760 \,$ps for an excited electronic state with a transition wavelength of about $950 \, $nm. This corresponds to a decay rate $\Gamma$ of about $1 \cdot 10^9 \,$Hz and a transition frequency $\omega_0$ of about $2 \cdot 10^{15} \,$Hz. Assuming moreover that $\Delta t = 3 \cdot 10^{-10} \,$s, we obtain a stationary state photon emission rate $I_{\rm ss}$ of about 8400 photons per second. Measuring the corresponding emission rate of about 4200 narrowband photons per second is experimentally feasible and suggests an experimental test of the validity of the master equation associated with the minimal coupling Hamiltonian \footnote{The reason why we consider a relatively small atom-detector distance of $10 \,$cm is to assure that $\Gamma \Delta t \ll 1$. This condition is required for the second order perturbation theory in Section \ref{more} to apply.}. 

Similar experimental tests of the multipolar and the minimal coupling Hamiltonians could be performed with chromium-based colour centers in diamond. Recently, M\"uller {\em et al.}  \cite{Jelezko} studied the photon emission of this system from an excited state with a lifetime of $1 \,$ns and a transition wavelength of $710 \,$nm at a temperature of $4\, $K. These parameters correspond to $\omega_0 = 2.6 \cdot 10^{15} \,$Hz and $\Gamma = 1 \cdot 10^9 \,$Hz. If we assume again a photon-absorbing environment which is $10 \,$cm or less away from the colour centre, the condition $\Gamma \Delta t \ll 1$ applies and the above calculations hold. In this case, the master equation associated with the minimal coupling Hamiltonian predicts a stationary state photon emission rate $I_{\rm ss}$ of about 2600 photons per second. 

\subsection{The rotating wave Hamiltonian}

Out of all the Hamiltonians considered in this paper, only the so-called rotating wave Hamiltonian predicts a stationary state photon emission rate 
\begin{eqnarray} \label{nomore}
I_{\rm ss} &\equiv & 0 \, . 
\end{eqnarray}
This suggests that this Hamiltonian identifies the components of the atom-field system correctly and yields the most accurate master equation. This master equation is the simplest example of the general master equation given in Eq.~(\ref{mstreq2}),
\begin{eqnarray}\label{mstreq4}
\dot{\rho}_{\rm A} &=& - {\rm i} \omega_0 \, [ \sigma_3 , \rho_{\rm A} ] +  {1 \over 2} A_- \left( 2 \sigma^- \rho_{\rm A} \sigma^+ - \sigma^+\sigma^- \rho_{\rm A} \right. \nonumber \\ 
&& \left. - \rho_{\rm A} \sigma^+\sigma^- \right) \, ,
\end{eqnarray}
since all the $u_k^+$'s vanish in this case (cf.~Table \ref{table1}). As one can see from Fig.~\ref{aminus}, the transition rate $A_-$ is essentially the same as the spontaneous decay rate $\Gamma$ in Eq.~(\ref{Gamma}). This means, Eq.~(\ref{mstreq4}) is essentially the same as the standard Born-Markov master equation which is usually obtained after applying the RWA and certain Markovian approximations. It is Markovian and of Lindblad form \cite{Lindblad}. As pointed out by Glauber \cite{Glauber}, there are hence realistic detectors which can implement the photon-absorbing measurements associated with this equation (cf.~Section \ref{discuss}).

\section{Conclusions} \label{conc}

This paper emphasizes that there are different ways of quantising the states of a two-level atom inside the free radiation field. Depending on the choice of gauge when writing down the classical Lagrangian, we obtain a different Hamiltonian $H$. On the quantum level, different Hamiltonians relate to each other via a unitary operator $R$ (cf.~Eq.~(\ref{h2})). All possible Hamiltonians $H$ are unitarily equivalent and possess exactly the same spectrum. The only difference is that each Hamiltonian refers to physically different components as representing the ``atom" and as representing the ``field." When deriving an atomic master equation, we assume a photon-absorbing environment and take the trace over the free radiation field. Each representation of the Hamiltonian hence results in a different master equation. 

In other words, every operator $R$ implies a different photon-absorbing environment and relates to a different detector model. The question we ask here is, which Hamiltonian identifies the free radiation field seen by a photon-absorbing environment correctly and therefore yields the most accurate master equation? This paper tries to answer this question without having to distinguish between a real and a virtual photon, whose definition depends on how the photon has been generated. Instead we derive master equations for a large set of unitary transformations $R_{\{\alpha_k\}}$ (cf.~Eq.~(\ref{ralphak})) which are able to generate the minimal coupling, the multipolar and the rotating wave Hamiltonians. We then propose to compare the predictions of the corresponding master equations with actual experimental findings.

Our derivation of atomic master equations avoids approximations whenever possible. The only approximations made in Section \ref{section3} are the assumption of an environmental response time $\Delta t$ and the use of second order perturbation theory with respect to an appropriately chosen interaction picture. The constant resetting of the free radiation field into its vacuum state is well justified in the presence of an external environment which selects this state via the absorption of photons (einselection) \cite{Zurek2}. This means, our model takes into account that photons fly away and do not return to interact again with their source. In the absence of an atomic source but the presence of a photon-absorbing environment, the vacuum state is the only state of the free radiation field which does not change in time. Moreover, second order perturbation theory is well justified as long as $\Delta t$ is short compared to the characteristic time scale of the effective atomic evolution. This paper avoids the RWA and takes the counter-rotating terms in the atom-field interaction Hamiltonian into account. The obtained general master equation contains transition rates $A_\pm$ and $B$ which depend explicitly on $\Delta t$ as well as on the cut-off frequencies $\omega_{\rm min}$ and $\omega_{\rm max}$ for the modes of the radiation field. A good estimate for $\Delta t$ is the light travel time to a detector or the walls of the laboratory. When the light travel time is relatively short, the time it takes a photon to be absorbed by the environment and the environment to re-thermalise should be added. 

In Section \ref{section4}, we calculate the stationary state photon emission rate $I_{\rm ss}$ in the absence of external driving for the master equations associated with the minimal coupling and the multipolar Hamiltonian. For the experimental parameters  of recent experiments with single quantum dots \cite{Atature} and chromium-based colour centers in diamond \cite{Jelezko}, we obtain enormous emission rates (cf.~Section \ref{figam}) which would have been noticed already. Out of all the atom-field Hamiltonians considered in this paper, the only one with $I_{\rm ss} \equiv 0$ (cf.~Eq.~(\ref{nomore})) is the so-called rotating wave Hamiltonian. Our analysis suggests that this Hamiltonian might yield the most accurate master equation. The name {\em rotating wave} derives from the fact that this Hamiltonian is almost the same as the atom-field Hamiltonian obtained after applying the RWA. However, the rotating wave Hamiltonian is not the result of an approximation but the result of a specific choice of the unitary transformation $R$ in Eq.~(\ref{h2}). 

When the Hamiltonian $H_{\rm I}$ in Eq.~(\ref{ham3}) contains counter-rotating terms, we obtain a non-zero stationary state photon emission rate $I_{\rm ss}$ which increases with decreasing $\Delta t$. This effect can be associated with the presence of a ``cloud" of virtual photons localised around the atom. A virtual photon is a photon generated by a term in the Hamiltonian which cannot conserve the free energy, ie.~a counter rotating term. When moving the photon-absorbing environment closer to the atom, the probability for the absorption of a virtual photon increases. Absorption of such a photon leaves the atom in an excited state, so that it may then decay via a (real) photon emission. In general, the $\{\alpha_k\}$ can be chosen to generate {\em any} value of $I_{\rm ss}$ whatsoever. In this way, it becomes possible for our model to reproduce detector dark count rates and finite temperature effects. \\[0.3cm]

\noindent {\em Acknowledgment.} A. B. thanks M. Atat\"ure and G. C. Hegerfeldt for helpful discussions. This work was supported by the UK Research Council EPSRC. 

\appendix
\section{Generalised PZW transformation} \label{app3}

In order to derive the general Hamiltonian $H' = R_{\{\alpha_k\}} \, H_{\rm min} \, R_{\{\alpha_k\}}^\dagger$, we use in the following the Baker-Campbell-Hausdorff formula 
\begin{eqnarray} \label{BCH}
{\rm e}^{X} Y {\rm e}^{- X} &=& \sum_{m=0}^\infty [ X,Y ]_m
\end{eqnarray}
with $[X,Y]_m = [X,[X,Y]_{m-1}]$ and $[ X,Y]_0 = Y$. As closer look at $H_{\rm min}$ in Eq.~(\ref{min3}) and $R_{\{\alpha_k\}}$ in Eq.~(\ref{ralphak}) shows that this calculation requires the commutators  
\begin{eqnarray} \label{commi}
\left[ {\bf A}_{\{\alpha_k\}}({\bf 0}) \cdot {\bf r}, {\bf p}^2 \right] &=& 2 {\rm i} \hbar \, {\bf A}_{\{\alpha_k\}}({\bf 0}) \cdot {\bf p} \, , \nonumber \\
\left[{\bf A}_{\{\alpha_k\}}({\bf 0}) \cdot {\bf r}, {\bf A}(0) \cdot {\bf p} \right] &=& {\rm i} \hbar \, {\bf A}_{\{\alpha_k\}}({\bf 0}) \cdot {\bf A}(0) \, , ~~~ \nonumber \\
\Big[{\bf A}_{\{\alpha_k\}}({\bf 0}) \cdot {\bf r} , \sum_{{\bf k}\lambda } \hbar \omega_k \, a_{{\bf k}\lambda }^\dagger a_{{\bf k}\lambda } \Big] &=&  {{\rm i} \hbar \over \epsilon_0} \, {\bf \Pi}_{\{\alpha_k\}} ({\bf 0}) \cdot {\bf r} \, , \nonumber \\
\left[{\bf A}_{\{\alpha_k\}}({\bf 0}) \cdot {\bf r}, {\bf A}_{\{\alpha_k\}}(0) \cdot {\bf p} \right] &=& {\rm i} \hbar \, |{\bf A}_{\{\alpha_k\}}({\bf 0})|^2 \, , \nonumber \\
\left[{\bf A}_{\{\alpha_k\}}({\bf 0}) \cdot {\bf r}, {\bf \Pi}_{\{\alpha_k\}}(0) \cdot {\bf r} \right] &=& \sum_{{\bf k}\lambda } {{\rm i} \hbar \over L^3} \, \alpha_k^2 \, |{\bf r}|^2 \, . ~~~
\end{eqnarray}
All other commutators which occur in the calculation of $H'$ equal zero. Using Eq.~(\ref{commi}), we obtain a Hamiltonian of the same form as the Hamiltonian in Eq.~(\ref{h1}). Its components can be found in Eq.~(\ref{ham}).

\section{Analysis of the transition rates $A_\pm$} \label{app}

The aim of this section is to determine the transition rates $A_\pm$ in Eq.~(\ref{b}) as functions of the environmental response time $\Delta t$ and of the typical cut-off frequencies $\omega_{\rm min}$ and $\omega_{\rm max}$ for the modes of the free radiation field. Using Eqs.~(\ref{u-2}) and (\ref{b}) and performing the summation over $\lambda$, one can show that 
\begin{eqnarray}
\sum_{{\bf k} \lambda} |g_{{\bf k}\lambda}|^2 = {e^2 \omega_0 \over 2 \epsilon_0 L^3 \hbar}  \, \sum_{{\bf k} \lambda} \left( 1 - |\hat {\bf d} \cdot \hat {\bf k}|^2\right) |{\bf d}|^2 \, , 
\end{eqnarray}
where $\hat {\bf d}$ and $\hat {\bf k}$ are unit vectors in the directions of ${\bf d}$ and ${\bf k}$, respectively. Here we are especially interested in the large volume limit, where $L \to \infty$. Next we choose a coordinate system such that ${\bf d}$ points in the $z$-direction and $\hat {\bf d} \cdot \hat {\bf k} = \cos \vartheta$ and replace the summation over ${\bf k}$ by an integration,
\begin{eqnarray}
\sum_{{\bf k} \lambda} \longrightarrow {L^3 \over 8 \pi^3 c^3} \, \int_{\omega_{\rm min}}^{\omega_{\rm max}} {\rm d} \omega_k \, \omega_k^2 \, \int_0^\pi {\rm d} \vartheta \, \sin \vartheta \int_0^{2 \pi} {\rm d} \varphi \, , ~~
\end{eqnarray}
where $c$ denotes the speed of light. Taking this into account, we then find that     
\begin{eqnarray} \label{rates}
A_\pm &=& {2 \Gamma \over \pi \omega_0^2 \Delta t} \int_{\omega_{\rm min}}^{\omega_{\rm max}} {\rm d} \omega_k \, f_\pm (\omega_k) \, \sin^2 \left[ {1 \over 2} (\omega_0 \pm \omega_k) \Delta t \right] \nonumber \\
\end{eqnarray}
with $f_\pm (\omega_k)$ defined as 
\begin{eqnarray} \label{rates2}
f_\pm (\omega_k) &\equiv & { \left(u_k^\pm \omega_k \right)^2 \over (\omega_0 \pm \omega_k)^2}  
\end{eqnarray}
and with the spontaneous decay rate $\Gamma$ defined as \footnote{When using the electric dipole and the RWA in the large volume limit for an environmental response time $\Delta t \to \infty$, the transition rates $A_\pm$ in Eq.~(\ref{mstreq2}) can be shown to equal  $A_+ = 0$ and $A_- = \Gamma$. This means, the spontaneous decay rate $\Gamma$ in Eq.~(\ref{Gamma}) is the usually obtained spontaneous decay rate for the excited atomic state $|1 \rangle$ \cite{Hegerfeldt93}. However, these approximations are not justified when using for example the multipolar and the minimal coupling Hamiltonians.}
\begin{eqnarray} \label{Gamma}
\Gamma &\equiv & {e^2 \omega_0^3 \over 3 \pi \epsilon_0 \hbar c^3} \, |{\bf d}|^2 \, . 
\end{eqnarray}
Table \ref{table1} summarises the $u_k^\pm$ and $f_\pm(\omega_k)$ for these three cases.

\begin{figure}[t]
\begin{minipage}{\columnwidth}
\begin{center}
\hspace*{-1.5cm} \includegraphics[scale=1]{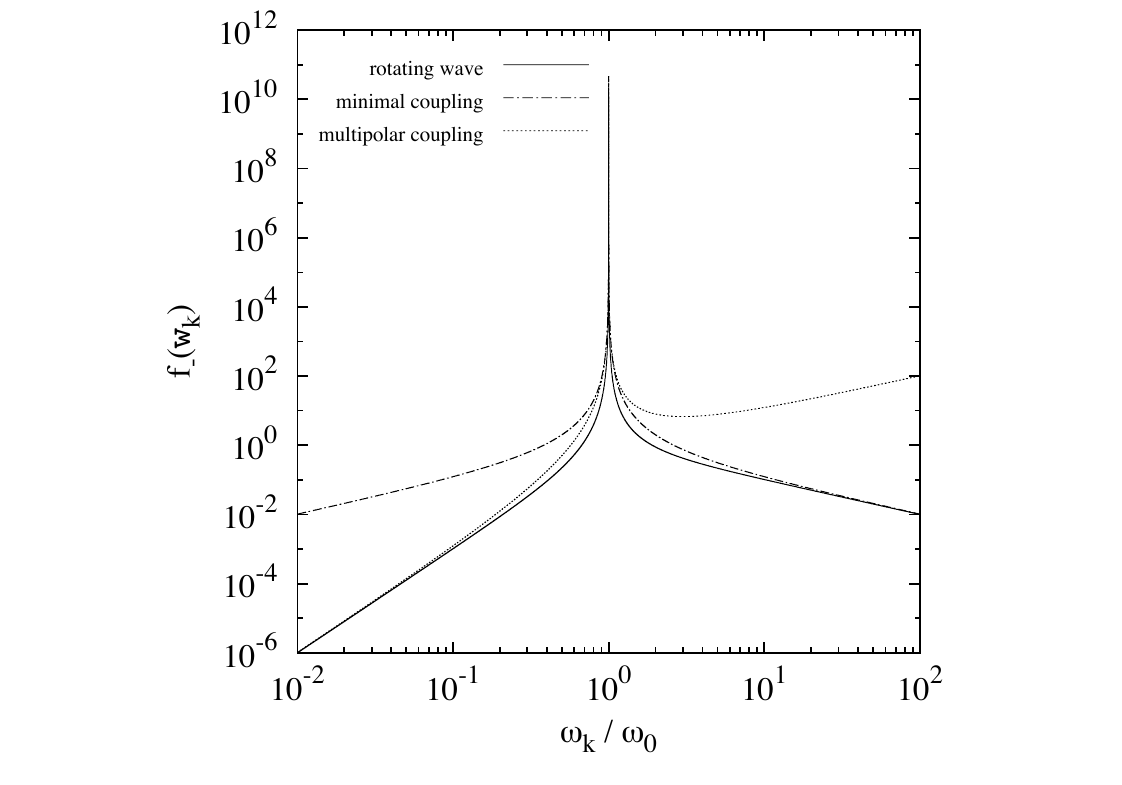}
\end{center}
\caption{Logarithmic plot of the function $f_- (\omega_k)$ in Eq.~(\ref{rates2}) for the rotating wave, the minimal coupling, and the multipolar Hamiltonian.} \label{fminus}
\end{minipage}
\end{figure}

Fig.~\ref{fminus} illustrates a removable singularity of the coefficients $f_- (\omega_k)$ at $\omega_k = \omega_0$. In case of the rotating wave and the minimal coupling Hamiltonian, $f_- (\omega_k)$ tends to zero when $\omega_k$ tends to infinity. This means, when performing the integration in Eq.~(\ref{rates}), we are likely to obtain a transition rate $A_-$ which depends only weakly on the cut-off frequencies $\omega_{\rm min}$ and $\omega_{\rm max}$ as long as both are sufficiently different from $\omega_0$. This is confirmed by Fig.~\ref{aminus} which shows that $A_-$ is in both cases essentially the same as the spontaneous decay rate $\Gamma$ in Eq.~(\ref{Gamma}). This applies for a wide range of environmental response times $\Delta t$ and a wide range of realistic cut-off frequencies $\omega_{\rm min}$ and $\omega_{\rm max}$. However, for the multipolar Hamiltonian, $f_- (\omega_k)$ grows rapidly when $\omega_k$ increases. It is therefore not surprising to see (cf.~Fig.~\ref{aminus}) that the corresponding transition rate $A_-$ in Eq.~(\ref{rates}) grows with $\omega_{\rm max}$.

\begin{table}[t]
\begin{tabular}{ccccc}  
        \hline \\[-0.2cm]
	~~~ $H$ ~~~ & ~~~ $u_k^-$ ~~~ & ~~~ $u_k^+$ ~~~ & ~~~ $f_-(\omega_k)$ ~~~ & ~~~ $f_+(\omega_k)$ ~~~
	\\[0.2cm] \hline \\[-0.2cm]
	$H_{\rm rot}$ & ${2 \sqrt{\omega_0 \omega_k} \over \omega_0 + \omega_k}$ & 0 & ${4 \omega_0  \omega_k^3 \over (\omega_0^2 - \omega_k^2)^2}$ & 0 
	\\ \\
	$H_{\rm min}$ & $\sqrt{\omega_0 \over \omega_k}$ & $\sqrt{\omega_0 \over \omega_k}$ & ${\omega_0 \omega_k \over (\omega_0 - \omega_k)^2}$ & ${\omega_0 \omega_k \over (\omega_0 + \omega_k)^2}$  
	\\ \\
	$H_{\rm mult}$ & $\sqrt{\omega_k \over \omega_0}$ & $- \sqrt{\omega_k \over \omega_0}$ & ${\omega_k^3 \over \omega_0 (\omega_0 - \omega_k)^2}$ & ${\omega_k^3 \over \omega_0 (\omega_0 + \omega_k)^2}$ 
	\\[0.4cm] \hline 
\end{tabular}
\caption{The coefficients $u_k^\pm$ in Eq.~(\ref{u-2}) and the coefficients $f_\pm(\omega_k)$ in Eq.~(\ref{rates2}) for the rotating wave, the minimal coupling, and the multipolar Hamiltonian.} \label{table1}
\end{table}

\begin{figure}[t]
\begin{minipage}{\columnwidth}
\begin{center}
\hspace*{-1.5cm} \includegraphics[scale=1]{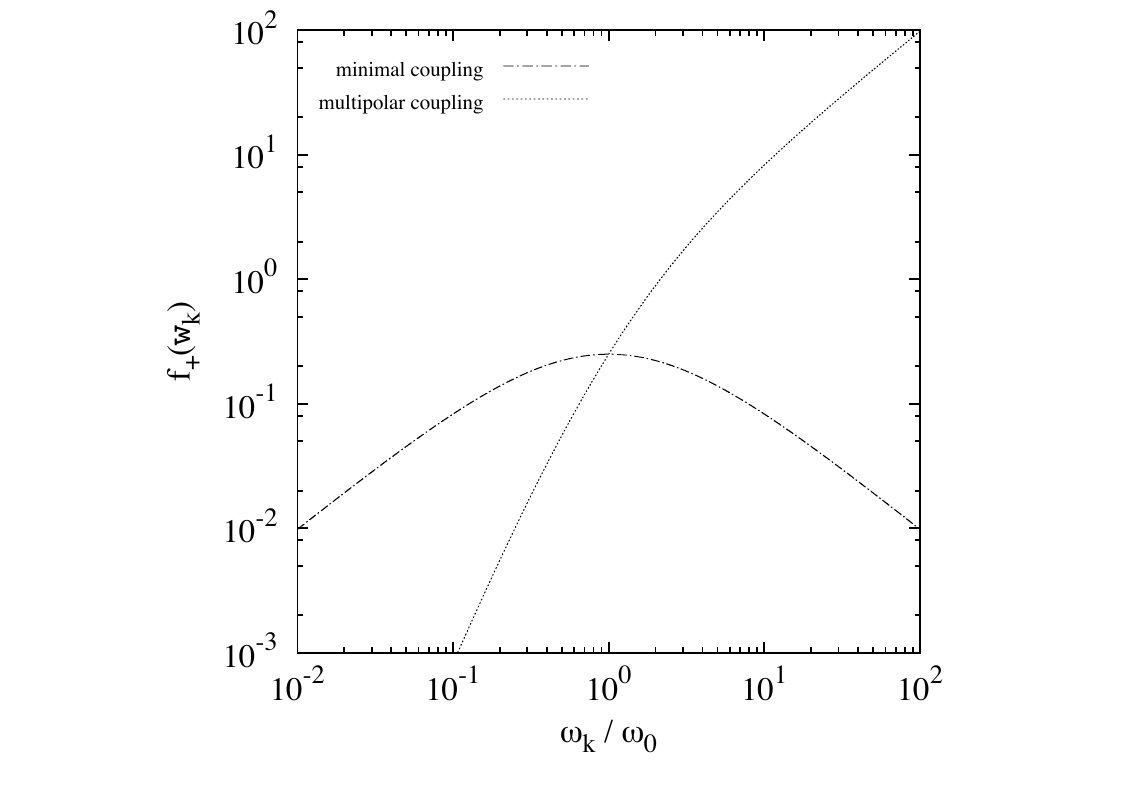}
\end{center}
\caption{Logarithmic plot of the function $f_+ (\omega_k)$ in Eq.~(\ref{rates2}) for the minimal coupling and the multipolar Hamiltonian. For the rotating wave Hamiltonian we have $f_+ \equiv 0$.} \label{fplus}
\end{minipage}
\end{figure}

Out of all the Hamiltonians considered in this paper, the rotating wave Hamiltonian is the only one for which the transition rate $A_+$ vanishes, ie.
\begin{eqnarray} \label{Aminu}
A_+ &=& 0 \, .
\end{eqnarray}
This applies, since $f_+(\omega_k) \equiv 0$ in this case, as Table \ref{table1} shows. The calculation of the transition rate $A_+$ for the minimal coupling and the multipolar Hamiltonians too is relatively straightforward, since $f_+ (\omega_k)$ has no singularity for finite values of $\omega_k$ (cf.~Fig.~\ref{fplus}). Relatively simple but also very accurate results are obtained, when replacing the $\sin^2$ function in Eq.~(\ref{rates}) by ${1 \over 2}$. This approximation is well justified, since $ \Delta t \gg 1/\omega_0$. Using this approximation and the values for $f_+(\omega_k)$ in Table \ref{table1}, we find that
\begin{eqnarray} \label{Aminx}
A_+ &=& {\Gamma \over \pi \omega_0 \Delta t} \left[ \, {1 \over x} + \ln x \, \right]_{\omega_{\rm min}/\omega_0+1}^{\omega_{\rm max}/\omega_0+1} ~~
\end{eqnarray}
in case of the minimal coupling Hamiltonian, while we obtain
\begin{equation} \label{Amin2}
A_+ = {\Gamma \over \pi \omega_0 \Delta t} \left[ \, {1 \over x} - 3 x + {1 \over 2} x^2 + 3 \ln x \, \right]_{\omega_{\rm min}/\omega_0+1}^{\omega_{\rm max}/\omega_0+1} 
\end{equation}
in case of the multipolar Hamiltonian. Both rates depend strongly on the environmental response time $\Delta t$ as well as on the cut-off frequencies $\omega_{\rm min}$ and $\omega_{\rm max}$. This is illustrated in Fig.~\ref{aplus} which shows $A_+$ as a function of $\omega_{\rm max}$ for a fixed $\Delta t$ and $\omega_{\rm min} = 0$.

\section{Lindblad form of the general master equation} \label{a1}

In this appendix we diagonalise the general master equation in Eq.~(\ref{mstreq2}) and identify necessary conditions for this master equation to be of Lindblad form \cite{Lindblad}. To do so, we notice that Eq.~(\ref{mstreq2}) is an equation in first standard form
\begin{eqnarray}\label{1stdform}
\dot{\rho}_{\rm A} &=& -{\rm i}\tilde{\omega}_0 \, [\sigma_3,\rho_{\rm A}] \nonumber \\ && +\sum_{n,m=1}^2 M_{nm} \left[ \tilde{\sigma}_n\rho_{\rm A}\tilde{\sigma}_m^\dagger - {1\over 2} \{ \tilde{\sigma}_m^\dagger \tilde{\sigma}_n,\rho_{\rm A} \} \right] \, , ~~~~
\end{eqnarray}
where $\{,\}$ denotes the anticommutator and where the $M_{nm}$ are the matrix elements of a Hermitian operator $M$. 
A transfer of this equation into Lindblad form, ie.~in the form
\begin{eqnarray}\label{diagme}
\dot{\rho}_{\rm A} &=& -{\rm i}\tilde{\omega}_0 \, [\sigma_3,\rho_{\rm A}]\nonumber \\ 
&&+\sum_{i=1}^2 \lambda_i \left[ L_i\rho_{\rm A}L_i^\dagger - {1\over 2} \{ L_i^\dagger L_i,\rho_{\rm A} \} \right] \, , 
\end{eqnarray}
can now be achieved via a diagonalisation of $M$. To do so, we write $M$ as 
\begin{eqnarray}
M &=& \sum_{i=1}^2 \lambda_i \, |\lambda_i \rangle \langle \lambda_i| \, ,
\end{eqnarray}
where the $\lambda_i$ and the $|\lambda_i \rangle$ are the eigenvalues and eigenvectors of $M$. One can now easily check that this matrix is diagonalised by the unitary matrix
\begin{eqnarray}
U &=& \sum_{i=1}^2 |e_i \rangle \langle \lambda_i| \, ,
\end{eqnarray}
where the $|e_i \rangle$ are the canonical vectors $|e_1 \rangle = (1,0)^{\rm T}$ and $|e_2 \rangle = (0,1)^{\rm T}$. The matrix $U$ is indeed such that  
\begin{eqnarray}
U^\dagger M U &=& \left( {\begin{array}{cc}
 \lambda_1 & 0  \\
 0 & \lambda_2  \\
 \end{array} } \right) \, .
\end{eqnarray}
Now, defining new operators $L_i$ such that
\begin{eqnarray}\label{L}
\left( \begin{array}{c} \tilde \sigma_1 \\ \tilde \sigma_2 \end{array} \right) &=& U^\dagger \left( \begin{array}{c} L_1 \\ L_2 \end{array} \right)  \, 
\end{eqnarray}
and substituting them into Eq.~(\ref{1stdform}) gives the diagonal master equation of Lindblad form in Eq.~(\ref{diagme}).

We now turn our attention again to the general master equation in Eq.~(\ref{mstreq2}). This master equation is of the same form as Eq. (\ref{1stdform}), if we define the operators $\tilde{\sigma}_i$ by $\tilde{\sigma}_1 = \sigma^+$ and $\tilde{\sigma}_2 = \sigma^-$, while defining the matrix $M$ by
\begin{eqnarray}
M &=& \left( {\begin{array}{cc} A_+ & B  \\ B^* & A_- \end{array} } \right) \, .
\end{eqnarray}
The corresponding Lindblad operators $L_i$ are always traceless. This means, the corresponding master equation (\ref{diagme}) is of Lindblad form, if the eigenvalues $\lambda_i$ are either positive or equal to zero. This applies when ${\rm det}(M) \ge 0$, ie.~when
\begin{eqnarray} \label{poscon}
A_+ A_- & \ge & |B|^2 \, .
\end{eqnarray}
The definitions of the coefficients $A_\pm$ and $B$ can be found in Eqs.~(\ref{b}) and (\ref{b2}). 

\begin{figure}[t]
\begin{minipage}{\columnwidth}
\begin{center}
\hspace*{-1.5cm} \includegraphics[scale=1]{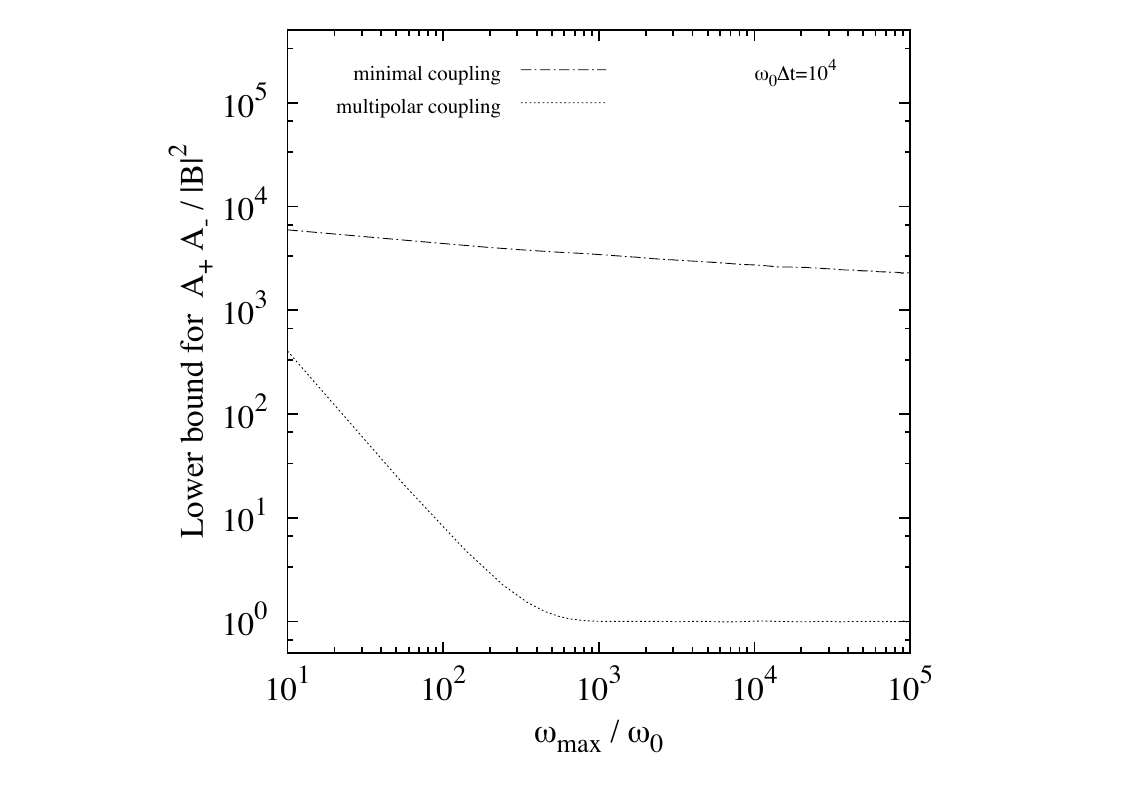}
\end{center}
\caption{Logarithmic plot of an upper bound for $A_+ A_- / |B|^2$ for the multipolar and the minimal coupling Hamiltonian obtained from a numerical integration of Eqs.~(\ref{rates}) and (\ref{b2}) after taking into account that $|d_1^2 + d_2^2 + d_3^2|^2 \le 1$ and that $\cos^2 (\omega_0 \Delta t) \le 1$, while assuming $\omega_{\rm min} = 0$. The figure confirms that Eq.~(\ref{poscon}) holds for the experimental parameters which we consider here and that the corresponding master equations are in general of Lindblad form.} \label{bbb}
\end{minipage}
\end{figure}

As it has been pointed out already in App.~\ref{app}, to calculate $A_\pm$ and $B$ a concrete representation of the atom-field Hamiltonian $H$ must be chosen. In the following, we proceed therefore as above and restrict ourselves again to the rotating wave, the minimal coupling, and the multipolar Hamiltonian. For example, in the case of the rotating wave Hamiltonian, we have $u_k^+ \equiv 0$ (cf.~Tab.~\ref{table1}) which implies $A_- = B = 0$. This means, one of the eigenvalues of $M$ is zero and the other one is positive. Eq.~(\ref{mstreq2}) is therefore of Lindblad form with one of the two Lindblad operators being zero. As Eq.~(\ref{mstreq4}) illustrates, the remaining non-zero Lindblad operator is the bare atomic lowering operator $\sigma^-$. 

To check whether Eq.~(\ref{poscon}) applies to the master equations associated with the multipolar and the minimal coupling Hamiltonian, we now proceed as in App.~\ref{app} and calculate $B$ as a function of $\Delta t$, $\omega_0$, $\omega_{\rm max}$, and $\Gamma$. To do so, we write the normalised atomic dipole moment $\hat {\bf d}$ and the general wave vector ${\bf k}$ as 
\begin{eqnarray}
{\bf d} = \left( \begin{array}{c} d_1 \\ d_2 \\ d_3 \end{array} \right) ~~ {\rm and} ~~ 
{\bf k} = \left( \begin{array}{c} \sin \vartheta \cos \varphi \\ \sin \vartheta \sin \varphi \\ \cos \vartheta \end{array} \right) \, ,
\end{eqnarray}
where $\vartheta$ and $\varphi$ are the usual polar coordinates for real vectors in three dimensions. Two normalised polarisation vectors $ {\rm{\bf e}}_{{\bf k} 1}$ and $ {\rm{\bf e}}_{{\bf k} 2}$ are then given by 
\begin{eqnarray}
{\rm{\bf e}}_{{\bf k} 1} = \left( \begin{array}{c} \cos \vartheta \cos \varphi \\ \cos \vartheta \sin \varphi \\ - \sin \vartheta \end{array} \right) \, , ~~ 
{\rm{\bf e}}_{{\bf k} 2} = \left( \begin{array}{c} - \sin \varphi \\ \cos \varphi \\ 0 \end{array} \right) \, .
\end{eqnarray}
One can easily check that these two vectors are pairwise orthogonal and orthogonal to ${\bf k}$. Substituting these into Eq.~(\ref{b2}) and performing the spatial integration over $\vartheta$ and $\varphi$, while taking the definition of the coupling constants $g_{{\bf k} \lambda}$ in Eq.~(\ref{u-2}) into account, we find that
\begin{eqnarray} \label{b3}
B &=& - {2 \Gamma \over \pi \omega_0^2 \Delta t} \, {\rm e}^{- {\rm i} \omega_0 \Delta t} \, \left( d_1^2 + d_2^2 + d_3^2 \right) \nonumber \\
&& \times \int_{\omega_{\rm min}}^{\omega_{\rm max}} {\rm d} \omega_k \, {\omega_k^2 \, u_k^+ u_k^- \over \omega_0^2 - \omega_k^2} \, \sin \left[ {1 \over 2} (\omega_0 + \omega_k ) \Delta t \right] \nonumber \\
&& \hspace*{2cm} \times \sin \left[ {1 \over 2} (\omega_0 - \omega_k ) \Delta t \right] \, .
\end{eqnarray}
Using trigonometric relations and approximating the integral over $\cos (\omega_k \Delta t)$ by zero, this expression simplifies to
\begin{eqnarray} \label{b4}
B &=& {\Gamma \over \pi \omega_0^2 \Delta t} \, {\rm e}^{- {\rm i} \omega_0 \Delta t} \, \cos (\omega_0 \Delta t) \, \left( d_1^2 + d_2^2 + d_3^2 \right) \nonumber \\
&& \times \int_{\omega_{\rm min}}^{\omega_{\rm max}} {\rm d} \omega_k \, {\omega_k^2 \, u_k^+ u_k^- \over \omega_0^2 - \omega_k^2} \, .
\end{eqnarray}
Concrete expressions for the $u_k^\pm$ can be found in Table~\ref{table1}. Replacing $|d_1^2 + d_2^2 + d_3^2|$ and $|\cos (\omega_0 \Delta t)|$ by their respective maximum of one, we obtain an upper bound for $|B|$. Fig.~\ref{bbb} uses this bound and verifies Eq.~(\ref{poscon}) for the multipolar and the minimal coupling Hamiltonian for a wide set of experimental parameters. This shows that the master equations associated with these two Hamiltonian are in general of Lindblad form.

\end{document}